\documentclass[twocolumn]{aastex631}

\usepackage{amsmath}
\usepackage{bm}

\begin{document}

\title{Magnetars in Binaries as the Engine of Actively Repeating Fast Radio Bursts}

\author[0000-0002-9725-2524]{Bing Zhang}
\affiliation{Hong Kong Institute for Astronomy and Astrophysics, University of Hong Kong, Pokfulam Road, Hong Kong, China}
\affiliation{Department of Physics, Department of Physics, University of Hong Kong, Pokfulam Road, Hong Kong, China}
\affiliation{Nevada Center for Astrophysics, University of Nevada, Las Vegas, NV 89154, USA}
\affiliation{Department of Physics and Astronomy, University of Nevada, Las Vegas, NV 89154, USA}
\email{bzhang1@hku.hk, bing.zhang@unlv.edu}

\author{Rui-Chong Hu}
\affiliation{Nevada Center for Astrophysics, University of Nevada, Las Vegas, NV 89154, USA}
\affiliation{Department of Physics and Astronomy, University of Nevada, Las Vegas, NV 89154, USA}
\email{ruichong.hu@unlv.edu}

\begin{abstract}
The association between FRB 20200428D and the Galactic magnetar SGR J1935+2154 makes magnetars the leading engine of cosmological fast radio bursts (FRBs). However, there is a list of puzzles for this magnetar-for-all-FRBs scenario: known Galactic magnetars are all isolated and none of them are active repeaters; some cosmological repeaters have extremely high repetition rates but without any measurable spin-related periodicity; some show long-term periodic, active windows; and some show diverse rotation measure (RM) evolution patterns, such as quasi-periodic fluctuations, sign reversals, and abrupt RM flares. Here we propose a unified theoretical paradigm for FRBs within the framework of the magnetar engine: Most active repeating FRBs originate from magnetars in binary systems with nearly aligned rotation and magnetic axes, some of which with a triple-aligned geometry, i.e. with an alignment with the orbital axis as well; whereas apparent non-repeaters and inactive repeaters originate from magnetars in isolated systems or in binaries with a misaligned geometry. By studying various magnetar formation channels using population syntheses, we show that a few percent of magnetars in the universe can be in binary systems, most with a massive star companion and some with aligned geometry. We suggest that such binary systems can account for the rich phenomenology of active repeaters. We suggest that the existence of a companion helps to maintain the aligned geometry and that the companion may play an active role in triggering FRBs in an active repeater source. 
\end{abstract}

\keywords{stars: magnetars --- binaries: general --- fast radio bursts }

\section{Introduction} \label{sec:intro}

Fast radio bursts (FRBs) are one of the most fascinating and mysterious objects in the universe \citep{lorimer07,thornton13,petroff19,zhang23}. Located at cosmological distances, they represent the most luminous radio bursts in the universe. The combination of high luminosity and short duration (typically milliseconds) makes them possess a brightness temperature of the order of $\sim 10^{36}$ K, which demands an extreme coherent radiation mechanism to power these events. The facts that some FRBs are repeaters \citep{spitler16,chime-new-repeaters} and that the event rate density is too high for catastrophic events \citep{ravi19b,luo20} suggest that the engine of most (if not all) FRBs should not be destructive but should be able to continuously sustain energy to power repeated bursts. In the literature, dozens of source candidates have been proposed \citep{platts19,zhang23}. 

The detection of FRB 20200428D from a Galactic magnetar SGR J1935+2154 \citep{CHIME-SGR,STARE2-SGR,HXMT-SGR,Integral-SGR} suggested that at least some FRBs originate from magnetars. The origin of cosmological FRBs, however, is still not fully solved. Two opposite views exist \citep[e.g.][]{zhang20b}: A conservative view suggests that all FRBs in the universe are powered by magnetars. A more open but speculative view, on the other hand, suggest that there are diverse origins of FRBs: besides repeaters, there are genuine non-repeaters of catastrophic origins; and even within repeaters, there are other types of sources besides magnetars. The current data cannot rule out either possibility. 

In this {\em Letter}, we discuss the former point of view, i.e. all the FRBs in the universe are powered by magnetars \citep[e.g.][]{lu20,margalit20,yangzhang21}. In Section \ref{sec:obs}, we discuss FRB observations and point out some puzzles and challenges related to this suggestion. In Section \ref{sec:magnetars}, we summarize various magnetar formation channels including those forming magnetars in single (\ref{sec:single}) or binary (\ref{sec:binary}) star systems and estimate their relative fractions (\ref{sec:fractions}). We then make the suggestion in Section \ref{sec:engines} that active repeating FRBs preferentially originate from magnetars in binaries with an aligned geometry and discuss how this scenario interprets various curious observational properties of active repeating FRBs (\ref{sec:alignment}-\ref{sec:triggers}). The conclusions are summarized in Section \ref{sec:conclusions}. 

\section{Puzzles and challenges to the magnetar model}\label{sec:obs}

Known magnetars in the Milky Way and nearby galaxies such as Large and Small Magellanic Clouds (LMC and SMC) (See the McGill Online Magnetar Catalog at: https://www.physics.mcgill.ca/$\sim$pulsar/magnetar
/main.html)  
are observed as soft gamma-ray repeaters (SGRs) or anomalous X-ray pulsars (AXPs) \citep{kaspi17}. They typically have a dipolar magnetic field of $10^{14}-10^{15}$ G and a spin period in the range of $1-12$ s. They are in isolated systems, and none of them have a known companion star\footnote{There are two marginal cases, i.e. SGR 0755-2933 and CXOU J171405.7-381031, that have faint infrared counterparts \citep{Doroshenko21,Richardson23,Chrimes2022}, but neither has been confirmed as a stellar companion.}. Only a small fraction appears as radio pulsars with clear periodicity identified from timing observations during certain periods of time \citep[e.g.][]{camilo07,zhu23}. Only one source, SGR J1913+2154, emitted one radio burst with a luminosity close to (but still about one order of magnitude smaller than) that of a typical cosmological FRB, together with more intermediate luminosity radio bursts \citep{CHIME-SGR,STARE2-SGR,kirsten21}. 

The engine of cosmological FRBs is more difficult to diagnose. Most CHIME-detected FRBs are one-off events, or apparent non-repeaters \citep{chime-1st-catalog}, whose radiation signals provide little clue regarding their origins. A fraction of FRBs repeat; some are active repeaters, which have been extensively studied with radio telescopes across the globe, such as CHIME \citep{chime-new-repeaters}, Arecibo \citep{scholz17}, and most extensively, China's FAST, which has the highest sensitivity \citep{lid21,xuh22,zhou22,zhangyk22,zhangyk23,Li2025,zhou25}. Some interesting features have been revealed, which are quite different from what is expected for an engine similar to known magnetars.

The magnetar-for-all-FRBs interpretation faces the following puzzles and challenges:

\begin{itemize}
\item The first puzzle is that after extensive searches, no credible spin-related periodicity has been identified from any of the active repeaters \citep{zhangy18,lid21,xuh22,niujr22,zhoudk25}. A detailed analysis of the radio pulses and radio bursts from the Galactic magnetar SGR J1935+2154 \citep{zhu23} suggested that unlike pulses that concentrate in fixed phases, bursts (FRBs and intermediate-luminosity bursts) typically occur more randomly in phase. In any case, for active repeaters, even if the bursts occur more randomly in phase, after detecting thousands of bursts in some of the sources, evidence of periodicty should have emerged, but this has not been the case. One speculation is that the spin and magnetic axes of the magnetars powering active repeaters may be nearly aligned \citep{beniamini25,luo25}. However, it remains a question why this is the case given that none of the Galactic magnetars show such a behavior. 
\item The second puzzle is that the immediate environment of active repeaters seems to be very different from that of Galactic magnetars. In particular, various observations are consistent with the hypothesis that the FRB engine resides in binary systems: 1. The CHIME-detected FRB 20180916B has a $\sim 16$-day periodicity, with detected bursts falling into a 5-day on-phase window \citep{chime-period}. The most straightforward interpretation is that the FRB engine is in a binary system, and the period is that of the orbital motion \citep[e.g.][]{ioka20}. Such an on/off, long-term periodicity is not commonly observed in other active repeaters, but a $\sim 160$-day period was later identified for FRB 20121102A with a wider active phase window \citep{rajwade20}. Other long-term periodicity was suggested in the rotation measure (RM) or dispersion measure (DM) evolution of some active repeaters, e.g. a possible $\sim 26$-d RM variation period of FRB 20201124A \citep{xujw25}, a possible $\sim 200$-d RM variation period of FRB 20220529 \citep{liangyf25}, a possible $\sim 126$-d DM variation period of FRB 20240209A \citep{pal25}, and a possible $\sim 361$-d DM variation period of GRB 20190520B (C.-H. Niu et al. 2025, submitted).
2. The rotation measure (RM) of some of these active repeaters (e.g. FRB 20201124A, FRB 20220529) do not stay constant. Rather, they show fluctuations \citep{xuh22,jiangjc22,Li2025} that are best interpreted as modulation of the magnetized companion \citep{wangfy22}. 3. RM reversals over the timescale of a few months have been observed in some actively repeating FRBs \citep[e.g.][]{anna-thomas23,Li2025}, and binary orbital motion is the most plausible interpretation \citep{zhang18b,yang23b}. 4. Finally, a significant ``RM flare'', i.e. a significant increase and decrease of RM by more than an order of magnitude within the timescale of weeks, was discovered in FRB 20220529 \citep{Li2025}. The most plausible interpretation is that a companion star ejected a clump of magnetized plasma similar to solar corona mass ejections, which passes across the line of sight, causing the ``RM flare'' signature. 
\item The third puzzle of FRB engines is that they have extremely high repetition rates. The most active Galactic magnetars emit a few tens of X-ray bursts (and a much smaller number of radio bursts) per hour 
\citep[e.g.][]{lin20}. Cosmological active repeaters, on the other hand, typically emit over a hundred FRBs per hour \citep{lid21,xuh22,zhangyk22,zhangyk23,Li2025,zhou25}. In particular, the latest active repeater FRB 20241114A had a maximum burst rate of $\sim 730 \ \rm hr^{-1}$, with a total of $>10,000$ bursts detected over a $>200$ day period with about 0.5-hour observing time per day \citep{zhangjs25}. It is hard to imagine how any of the magnetar burst triggering mechanisms \citep[e.g.][]{wangwy24} can manage to sustain such a high burst rate over such a long period of time. 
\end{itemize}

These puzzles naturally raise the following questions: If the engine of cosmological FRBs is indeed magnetars, what makes the difference between cosmological magnetars and Galactic magnetars? Are they formed from different progenitor channels? How do different channels generate magnetars with such different properties?

\section{Magnetar formation channels}\label{sec:magnetars}

In this section, we review different magnetar formation channels (see R.-C. Hu \& B. Zhang, 2025, in prep, for a detailed population synthesis analysis), with the focus on the fraction of magnetars in binary systems and the properties of those magnetars. 

\subsection{Summary flowchart}\label{sec:flowchart}

Magnetars can be formed from various progenitor channels involving massive stars in single or binary systems through dynamo mechanisms, tidal interactions, or mergers of degenerate stars such as white dwarfs and neutron stars. The final magnetar product can be either in an isolated star system or in a binary system with a companion. The bottom line is that the progenitor star needs to possess a large enough magnetic field seed at the stellar core \citep{ferrario06,spruit08} 
or to attain a large angular momentum at the core for a dynamo mechanism to operate. One dynamo mechanism discussed in the early literature was the $\alpha-\Omega$ dynamo, which can operate when the proto-neutron star (PNS) spins with the nearly maximum angular velocity \citep{duncan92,thompson93}. Such a dynamo mechanism tends to produce millisecond magnetars at birth. Observations of the associated supernova remnants (SNRs) of some magnetars suggest the total kinetic energy of the SNRs is at least one order of magnitude smaller than that predicted by a millisecond magnetar progenitor \citep{vink06}. Also, the properties of magnetar-associated SNRs are not that different from those associated with other neutron stars \citep{martin14}. This suggests that other less demanding dynamo mechanisms may be at play at the birth of most magnetars. One attractive mechanism is the Tayler-Spruit dynamo \citep{taylor73,spruit02}, which invokes a constructive feedback loop between differential rotation and winding from poloidal to toroidal magnetic field and an internal magnetic instability. Such a dynamo process is expected to operate during fall-back accretion of a newly formed proto-neutron star (PNS) after a core-collapse supernova explosion or a stellar merger. Magnetars can be generated with an initial spin period of $\lesssim 8$ ms \citep{Barrere2022}, which satisfies the observational constraint. In the following discussion, the Tayler-Spruit dynamo will be regarded as the main dynamo mechanism to form magnetars. 

An important property of the magnetar engine of FRBs is the inclination angle between the rotation and magnetic axes, $\chi$. As discussed in \cite{beniamini25} and \cite{luo25}, active repeaters likely require a nearly aligned geometry. In our study, we pay special attention to whether the formed magnetars have an aligned geometry. The detailed discussion is presented in Section \ref{sec:alignment}.

\begin{figure*}[p]
\centering
\includegraphics[width=0.95\linewidth, trim = 0 0 0 0, clip]{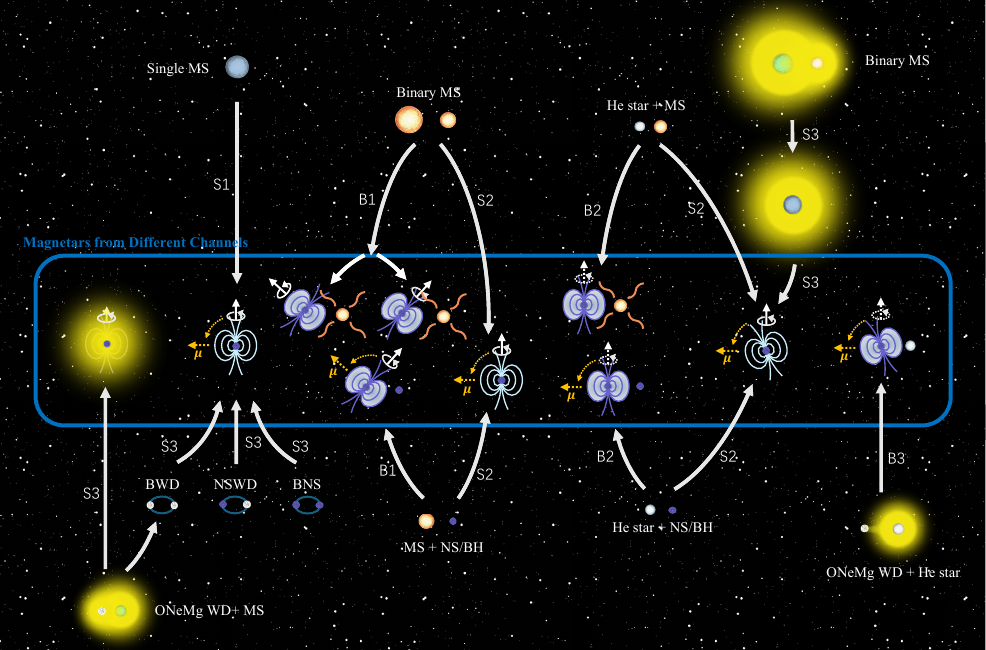}
\caption{Various magnetar formation channels through single and binary evolution. Magnetars with a black-colored magnetic field background are isolated, while those with a light-colored background are in binary systems. The figure also shows the magnetar's spin axis (white) and magnetic axis (orange, when mis-aligned from the spin axis), with the curved orange arrow indicating the flip evolution of the magnetic axis. The dashed lines indicate that, in some cases, the spin and magnetic axes of magnetars differ from what the flowchart shows.}
\label{fig:flowchart}
\end{figure*}

Various possible channels are summarized in Figure \ref{fig:flowchart}. Those marked as ``S'' and ``B'' denote the final magnetar being in single and binary systems, respectively. Within each type, the channels leading to these products are labeled numerically. In the following, we discuss these formation channels in more detail.

\subsection{Channels forming single magnetars}\label{sec:single}
In general, there are three channels to form magnetars in isolated systems:
\begin{itemize}
    \item Single-Star Evolution (S1):
            A magnetar can be in principle formed from a single-star progenitor after a supernova explosion if the star possesses a large seed magnetic field or a high spin. The required seed $B$ field or spin are quite demanding which can be satisfied by only a very small fraction of stars. A more practical and common channel is that a magnetar is formed in a slowly rotating single star, but the PNS after core collapse is spun up by supernova fallback accretion. The Tayler-Spruit dynamo then amplifies the large-scale magnetic field, leading to the formation of a magnetar \citep{Barrere2022}. In the following, we assume that the majority of magnetars in the S1 channel are formed via the Tayler-Spruit dynamo mechanism.
    \item Binary Disruption (S2):
            In this channel, the progenitor of the magnetar was in a binary system. However, the magnetar received a large enough kick during the supernova explosion, so the binary system was disrupted, leaving behind an isolated magnetar \citep{Chrimes2022, Sherman2024}. The formation of the magnetar before disruption may be similar to the fall-back accretion / Tayler-Spruit dynamo mechanism as discussed for the S1 channel. Alternatively, the magnetar could also be formed via He star - main sequence (MS) star interactions as discussed in B2, but with a large enough kick during a core collapse supernova explosion of the He star to disrupt the progenitor binary system. 
    \item Stellar Merger (S3): 
    An isolated magnetar can also be formed through various stellar merger processes: \begin{itemize}
                \item 
                First, one can have a merger of binary main-sequence (MS) stars before either star goes through a supernova explosion. Such a merger may produce a massive star with a strong, large-scale magnetic field. The explosion of such a merged star could ultimately make a magnetar after a core-collapse supernova (CCSN) explosion \citep{Schneider2019,Shenar2023,Frost2024}. 
                \item Alternatively, if a merger occurs during the common envelope (CE) phase of a magnetized ONeMg white dwarf and the core of a hydrogen-rich or helium-rich nondegenerate star, it can result in the formation of a magnetar surrounded by a massive envelope \citep{Ablimit2022b}.
                \item Finally, magnetars can also be formed through mergers of compact objects (CO), including binary white dwarfs (WD-WD) \citep{Schwab2021}, a WD-NS system \citep{Zhong2020,yangj22}, or even a binary neutron star (NS-NS) system \citep{Reboul-Salze2025}\footnote{The possibility that NS-NS mergers can make long-lived magnetars has long been discussed within the context of gamma-ray bursts \citep[e.g.][]{dai06,gaofan06}, electromagnetic counterparts of gravitational waves \citep[e.g.][]{zhang13,gao13,yu13,gao16,ai21}, and FRBs \citep[e.g.][]{margalit19,wang20}.}.
        \end{itemize}
\end{itemize}

\subsection{Channels forming magnetars in binaries}\label{sec:binary}

A small fraction of magnetars can still reside in binary systems after birth. The possible formation channels include the following:

\begin{itemize}
    \item Binary Surviving SN Explosion (B1): A progenitor similar to the S2 channel can produce a magnetar in a binary system if the system remains bound after the SN explosion. The orbit tends to have a large eccentricity because of the large kick received by the magnetar during the SN explosion. In this channel, the magnetar is formed through fallback accretion from the stellar envelope or supernova ejecta via the Tayler-Spruit dynamo mechanism.  
    \item Tidal Spin-Up (B2):
    This is a special channel proposed by \cite{Hu2023} that invokes a helium star - MS star binary in a tight orbit. The He star is spun up through tidal interactions by its companion, after a common envelope or stable mass-transfer phase. This enhanced rotation leads to a rapidly spinning core, which can make fast-spinning magnetars without the need of a fall-back-induced Tayler-Spruit dynamo process. Because the He star is envelope-stripped, the supernova explosion would proceed with a smaller kick than a regular CCSN, so that the chance of survival of the binary system is much higher.
    \item Accretion-Induced Collapse (B3): Another channel that can form magnetars in a binary system without significant kick at the birth of magnetar is through the accretion-induced collapse (AIC) of an ONeMg white dwarf \citep{Ablimit2022a}. The accreted mass comes from a low-mass companion. This channel is not expected to be associated with a supernova, and no significant kick is expected. 
\end{itemize}

\subsection{Fractional estimates}
\label{sec:fractions}

\begin{figure}[t]
\centering
\includegraphics[width=1.0\linewidth, trim = 0 0 0 0, clip]{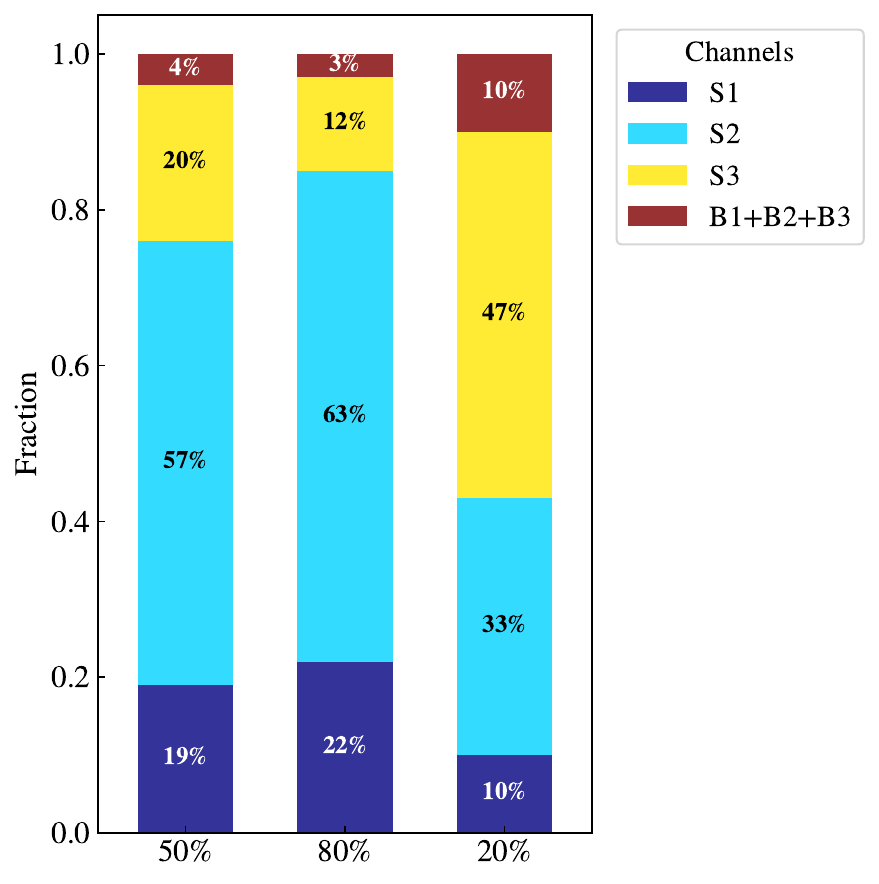}
\caption{Fractions of magnetars from different formation channels based on the fiducial model. Different bars represent different magnetar formation fractions among all NSs.}
\label{fig:magnetar_fraction}
\end{figure}

A detailed study of the magnetar formation channels through population syntheses is presented in our companion paper (R.-C. Hu \& B. Zhang, 2025, in prep). Here we present some related results from the models discussed in that paper. Due to the significant uncertainties in the amount of fallback accretion during core collapse supernovae (CCSNe), which affects whether the Tayler–Spruit dynamo can be triggered to produce strong magnetic fields, the formation fraction of magnetars among all neutron stars is taken as a free parameter to be constrained from observations. We adopt three values 20\%, 50\%, and 80\%, to cover a wide range of possibilities \citep{Beniamini2019}, but take 50\% as the fiducial value because this fraction seems to be supported by the observational data (N. Rea, 2025, private communication). The results of the contributions of different magnetar formation channels are presented in Figure \ref{fig:magnetar_fraction} and Table \ref{tab:Fractions}. 
The results show that only a few percent of newly-formed magnetars remain in binary systems. The majority of magnetars are isolated systems ($\sim 96\%$ for the fiducial model), including those originating from single-star evolution (S1, $\sim 22\%$ for the fiducial model), SN-disrupted binaries (S2, $\sim 57\%$ for the fiducial model), and binary mergers (S3, $\sim 19\%$ for the fiducial model), respectively. This dominance of isolated magnetars is in general consistent with observational data of Galactic magnetars, which show no firm evidence of the existence of a companion star \citep[e.g.][]{Olausen2014}.
 
As shown in Figure \ref{fig:magnetar_fraction}, increasing the assumed magnetar-to-NS formation fraction leads to a higher inferred rate of magnetars formed via the Tayler–Spruit dynamo, thereby increasing the contributions from the S1 and S2 channels. However, when the assumed magnetar formation fraction is too low, the contribution from CCSNe is reduced, and the merger channel (S3) becomes the dominant one for magnetar formation. In this case, the fraction of magnetars residing in binary systems also increases, potentially conflicting with the observational data.

As shown in Table \ref{tab:Fractions}, we also examine the contributions from various sub-channels. In the S2 channel, binary disruption can result from different types of supernovae, each imparting distinct natal kicks. In our fiducial model, the majority of magnetars in the S2 channel originate from the systems disrupted by strong kicks associated with CCSNe. For the S3 channel, we explore several different scenarios, with the dominant contribution arising from mergers between binary MS stars.

\begin{deluxetable}{ccccc}[t]
\tablewidth{\textwidth}
\tablecaption{Fractions of Various Magnetar Formation Channels for Different Models} \label{tab:Fractions}
\tablehead{
  \colhead{Channel} & \colhead{Sub-Channel} & \colhead{Fiducial (50\%)} & \colhead{High (80\%)} & \colhead{Low (20\%)}
}
\startdata
$\mathrm{S}1$   & ---     & $19\%$ & $22\%$ & $10\%$ \\
$\mathrm{S}2$   & ---     & $57\%$ & $63\%$ & $33\%$ \\
$\mathrm{S}3$   & BMS     & $19\%$ & $11\%$ & $45\%$ \\
$\mathrm{S}3$   & CMIC    & $\sim 0.1\%$ & $\sim 0.1\%$ & $\sim 0.1\%$ \\
$\mathrm{S}3$   & BCO     & $1\%$ & $1\%$ & $2\%$ \\
$\mathrm{B}1$   & ---     & $1\%$ & $1\%$ & $3\%$ \\
$\mathrm{B}2$   & ---     & $3\%$ & $2\%$ & $7\%$ \\
$\mathrm{B}3$   & ---     & $\sim 0.1\%$ & $\sim 0.1\%$ & $\sim 0.1\%$ \\
\enddata
\tablecomments{The columns are (1) the formation channels; (2) sub-channels of S3 (BMS: binary main-sequence stars, CMIC: core merger induced collapse, BCO: binary compact objects); (3-5) fractions of each channel contributing to the total magnetar population, different columns represent different magnetar formation fractions among all NSs (50\%, 80\%, 20\%).}
\end{deluxetable}

\begin{figure*}
\centering
\includegraphics[width=0.60\linewidth, trim = 0 0 0 0, clip]{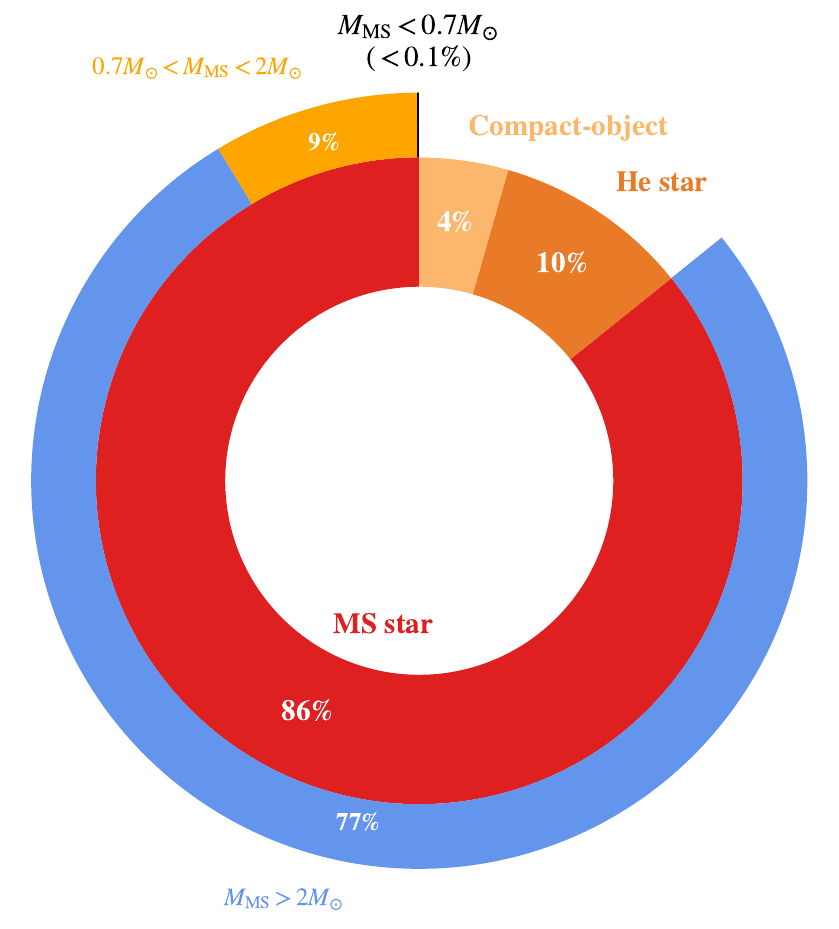}
\caption{Fractions of different magnetar companion types, including MS star subtypes.}
\label{fig:comp_type}
\end{figure*}

\subsection{Companion types}
\label{sec:companion}

For magnetars in binary systems, we also look into the details of the companion types. Figure \ref{fig:comp_type} displays the companion types for our population synthesis fiducial model. One can see that the majority of the companions (86\%) are MS stars, with 77\% having $M_{\rm MS} > 2 M_\odot$ and another $\sim 10\%$ having He star companions. The fraction with a compact star (WD or NS) companion is very small ($\sim 4\%$). One can see that overall the chance of having a massive star companion is very high. This is consistent with the conclusion of a previous investigation \citep{zhanggao20}.

\section{Magnetars in binaries as the engine of actively repeating FRBs}\label{sec:engines}

In this section, we argue that a good fraction of magnetars in binary systems tend to make nearly aligned spin and magnetic axes ($\chi \sim 0$), so are likely the engines of actively repeating FRB sources. We first discuss the channels that form magnetars with such alignment (sometimes triple alignment) and the subsequent $\chi$-evolution (\S\ref{sec:alignment}). We then discuss how such systems can produce the diverse observed evolution patterns of the RM as observed (\S\ref{sec:RM}). We also show that triple-aligned systems facilitate the propagation of FRBs generated from inside the magnetar magnetospheres (\S\ref{sec:propagation}). In \S\ref{sec:triggers}, we speculate that the existence of a companion may intrinsically enhance the burst triggering rate through perturbation of the magnetar magnetosphere by the companion winds. {\bf Finally, we estimate the ratio of active repeaters to non-repeaters within the proposed paradigm in \S\ref{sec:ratio}.}

\subsection{Spin-magnetic-axes  alignment}\label{sec:alignment}

The first puzzle of active repeaters, i.e. the lack of detectable periodicity after extensive searches \citep{zhangy18,lid21,xuh22,niujr22,zhoudk25}, can be most conveniently interpreted as the nearly aligned geometry between the magnetic and rotational axes (i.e. $\chi \sim 0$) of a magnetar engine. As is well known in the pulsar community, such a configuration tends to produce pulses with a large duty cycle. When the inclination angle $\chi$ approaches 0, the duty cycle is close to 100\%, and hence, no periodicity can be measured. Within the framework of FRBs, this possibility was independently suggested by \cite{beniamini25} through theoretical arguments and by \cite{luo25} through detailed modeling of the waiting time distribution and the energy distribution of the bursts from a few FAST-monitored active repeaters. Note that such an argument requires that FRB emission originates from the open field line region of a magnetar \citep[e.g.][]{kumar17,yangzhang18,wang19,kumar20a,lu20,yangzhang21,zhang22,qu22b,quzhang24}. The magnetospheric origin of the FRB emission has been overwhelmingly supported by the observational data \citep{luo20,niujr24,jiangjc24,mckinven25,nimmo25}.

Reviewing the magnetar formation mechanisms, we expect that the following channels make aligned spin and magnetic axes at birth:
\begin{itemize}
    \item All the mechanisms invoking differential rotation induced dynamo mechanisms (e.g. Tayler-Spruit dynamo);
    \item Magnetars formed through tidal spin-up of the progenitor star but the system did not go through violent core collapse supernova phase with large kicks.
\end{itemize}
Misaligned magnetars may be formed via CCSN explosion without significant Tayler-Spruit dynamo operation (e.g. in the S3 channel that invokes the merger of a binary MS system to produce a rapidly spinning core). This is because a CCSN is usually associated with a kick of the newborn PNS. Even though the origin of the kick is subject to debate, most of these mechanisms invoke a misaligned configuration in asymmetric explosions, with some predicting the alignment between the spin axis and the kick velocity \citep[e.g.][and references therein]{janka12,beniamini24}.

As a result, we expect that essentially all magnetar formation channels, both for binary (B1, B2, and B3) and single channels (S1 and S2) will form aligned magnetars ($\chi \sim 0$) at birth. Whether an FRB magnetar engine has an aligned geometry depends on the subsequent evolution of the inclination angle $\chi$ after the magnetar is formed. 

The $\chi$ evolution of an isolated magnetar is known to be decided by the competition between spin-down-induced decreasing trend and viscosity-induced increasing trend (see Appendix \ref{sec:A-iso}). The pulsar spin-down torque tends to reduce $\chi$, and hence, maintain $\chi \sim 0$ for an initially nearly aligned rotator. On the other hand, a magnetar with a dominant toroidal magnetic field is deformed into a prolate shape. When there is a small mis-alignment between the spin and magnetic axes, a newborn, internally hot, star would undergo free precession, giving rise to an internal viscous dissipation torque that tends to induce ``spin-flip'' to make a perpendicular rotator ($\chi \sim 90^\circ$) with an oblate configuration \citep[e.g.][]{Lander2018,cutler02,lasky16}. This process requires the internal temperature of the magnetar to exceed a critical value \citep{lasky16}. Otherwise, $\chi$ will freeze to the initial configuration $\chi=0$. The timescale for such a flip, if it happens, depends on the complicated physical conditions such as the strength of the magnetic field, angular velocity, and temperature of the PNS, which can range from hours to millions of years. In any case, the fact that at least some Galactic magnetars do not show aligned geometry \citep[e.g.][]{zhu23} suggests that $\chi$ evolution is probably common for isolated magnetars.

The $\chi$-evolution of magnetars in binary systems is more complicated because there are other external torques, e.g. those related to tidal or magnetic interactions or directly from accretion of the wind of the companion, to affect the $\chi$-evolution. If the long-term period (e.g. the 16-day period for FRB 20180916B) is the orbital period, the tidal deformation due to the companion star would be too small compared to the magnetic deformation of the magnetars \citep[e.g.][]{lai93,lai94}. A more plausible mechanism is to influence the $\chi$-evolution through physical mass transfer, especially through wind accretion. It has been shown by \cite{biryukov21} that for a wind-fed high-mass X-ray binary, the random change of the spin-up torque makes $\chi$ go through a random walk evolution, with two attractors at $\chi=0$ and $\chi=90^{\circ}$. However, for an initial $\chi \sim 0$ configuration, especially for the triple-aligned configuration expected in a good fraction of systems, the probability of the final $\chi \sim 0$ configuration is greatly enhanced (see Appendix \ref{sec:A-bin}). We therefore expect that for magnetars in binary systems, the chance of having $\chi \sim 0$ for an extended period of time is much higher than that of isolated magnetars. The fact that active repeaters emit highly coherent radio emission from polar cap regions suggests that accretion has largely ceased when FRB emission is turned on, but an earlier accretion phase could have stabilized the inclination angle $\chi$ and suppressed the potential spin-flip. The companion is likely still ejecting an unsteady wind, which would carry magnetized materials along the line of sight. This is consistent with the erratic RM-evolution behavior observed in many active repeaters (see \S\ref{sec:RM} for more discussion). 

As shown in Figure \ref{fig:flowchart} and Table \ref{tab:Fractions}, a good fraction of binaries (those from the B2 channel, which has a higher fraction than B1 and B3 combined) not only possess nearly aligned magnetic and spin axes, the orbital angular momentum is also aligned along a similar direction. Therefore, such triple-aligned systems are common among our suggested FRB sources. There are exceptions of course. For example, if the magnetar is born from a violent CCSN with a large kick (B1), it is likely that the kick velocity vector (also the spin axis under some kick models) is in the orbital plane. The Tayler-Spruit dynamo would collect SN ejecta in the kick velocity direction, which produces a magnetar with nearly aligned magnetic and spin axes ($\chi \sim 0$) lying in the orbital plane \citep{Janka2022}. However, such a configuration may not be stable. The wind torque does not always help to maintain alignment, and the magnetar would likely undergo spin-flip similar to isolated magnetars. As a result, such binary systems likely make nearly orthogonal rotators with a companion, which may appear as inactive repeaters and observed as one-off FRBs.

\subsection{Diverse RM evolution behaviors}\label{sec:RM}

\begin{figure*}[htbp]
    \centering
    \includegraphics[width=\textwidth]{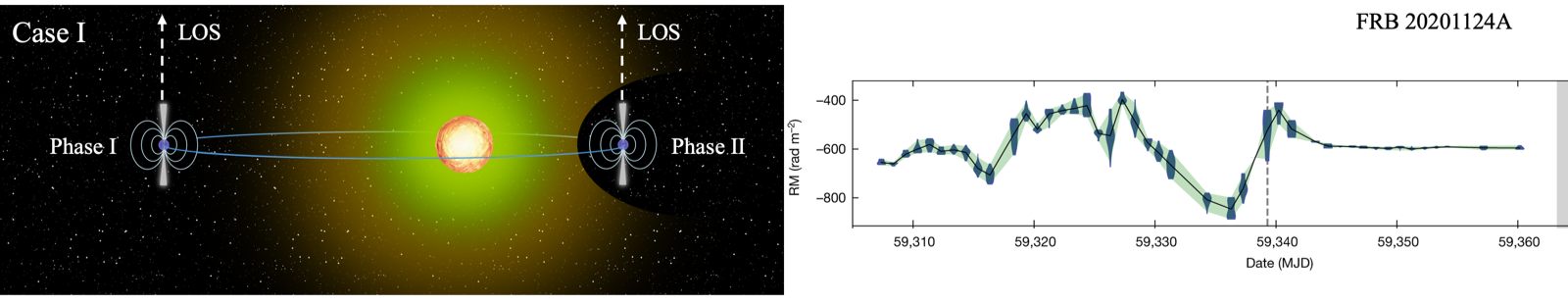}\hfill
    \includegraphics[width=\textwidth]{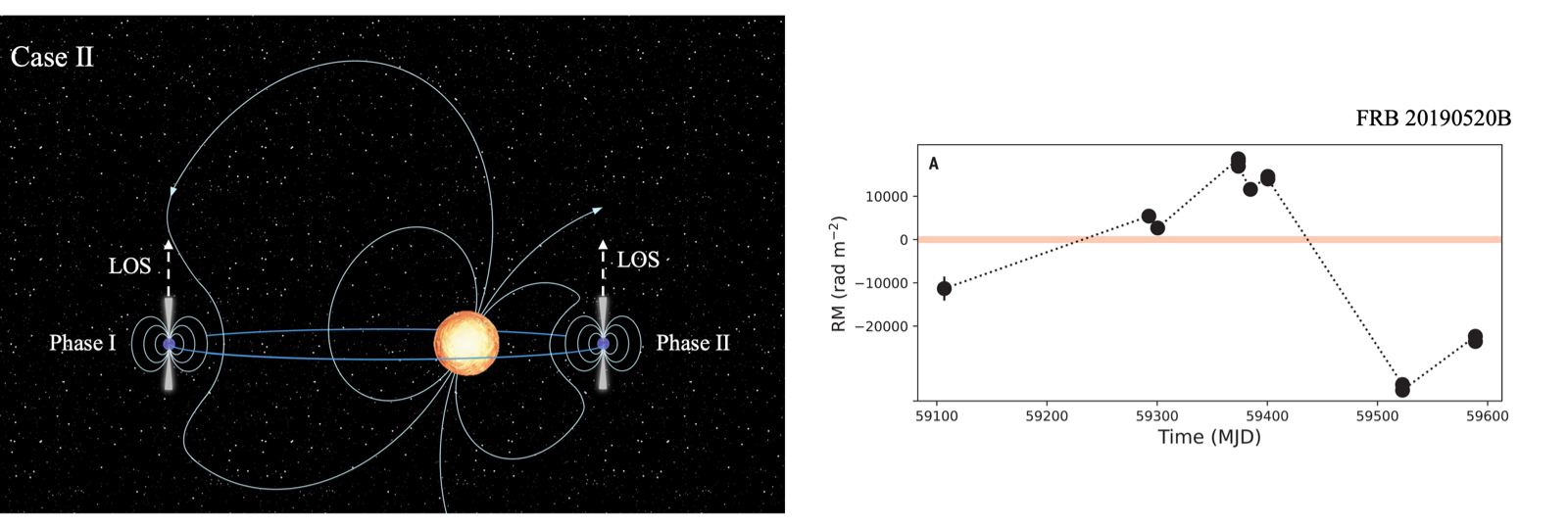}\hfill
    \includegraphics[width=\textwidth]{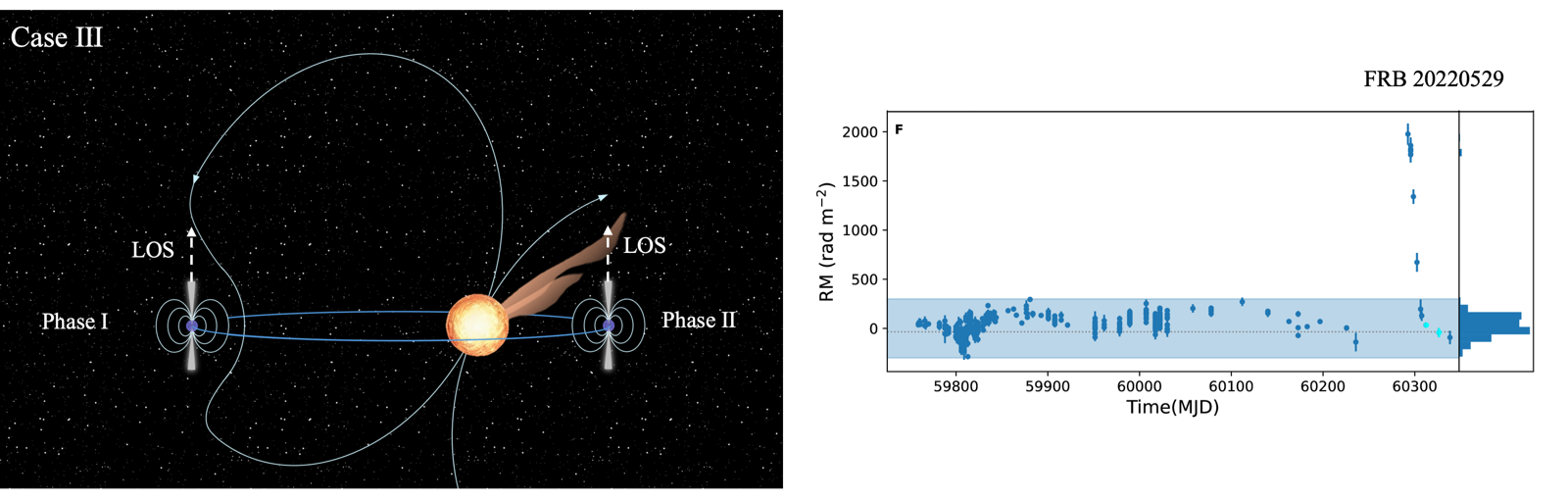}

    \caption{Illustration of three possible scenarios explaining the observed RM evolution in different FRBs. In each case, the left panel presents a cartoon showing different phases of a binary system and the surrounding environment of magnetars, with blue solid lines representing the orbits, white solid lines representing magnetic field lines, and white dashed lines representing the line of sight (LOS). The right panel shows the corresponding RM evolution over time (Case I: Irregular RM evolution of FRB 20201124A \citep{xuh22}; Case II: RM reversals of FRB 20190520B \citep{Anna2023}; Case III: RM flare and RM reversals in FRB 20220529 \citep{Li2025}).}
    \label{fig:three_figures}
\end{figure*}

Can an aligned magnetar binary system with a massive star companion and an unsteady stellar wind reproduce the observational properties of most active repeaters? We argue that the answer is ``yes''. In the following, we discuss several cases and show how the proposed system may account for the various RM evolution behaviors as observed. Figure \ref{fig:three_figures} shows the geometric configurations envisaged that may account for the data. Note that we have adopted a triple-aligned configuration for the illustration purpose, but a smaller fraction of aligned magnetar binary systems can have the orbital direction not aligned with the magnetic/spin axes of the magnetar (e.g. the B1 channels).

\begin{itemize}
    \item Irregular RM evolution patterns: Case I presented in Fig.\ref{fig:three_figures} addresses the irregular RM evolution as observed in the first active episode of FRB 20201124A \citep{xuh22}. This source had three active phases, as observed with the FAST telescope \citep{xuh22,zhou22,zhangyk22,xujw25}. The RM evolution during the first episode \citep{xuh22} is presented in the right panel of Fig.\ref{fig:three_figures} Case I. In particular, the RM evolution during the first 34 days is very significant, but during the last 20 days it remains nearly constant. Within our model, the sudden disappearance of RM variability can be interpreted as one of the following two possibilities: 1. The orbit has a large eccentricity. The environment is messy near the periastron, so a large RM variation is observed. Near the apastron, the environment is much cleaner so RM variation is stopped. 2. The early variation phase corresponds to the epoch when the companion is actively ejecting winds toward the FRB engine, and the late no-variation phase corresponds to the epoch when the companion stops wind ejection or eject winds in the directions not intersecting with the line of sight. A combined analysis of the RM data during all three active epochs of FRB 20201124A \citep{xujw25} revealed a possible $\sim 26$-day period in RM variation. If this is the orbital period, then the second option above maybe the more probably interpretation. 
    \item RM reversals: Some active repeaters, such as FRB 20190520B \citep{niu22,anna-thomas23} and FRB 20220529 \citep{Li2025}, have shown RM sign reversals over a timescale of several months. This marks the change in the projected global magnetic configuration, which signifies a binary system with a magnetized companion, especially if the emission beam is close to the orbital plane \citep[e.g.][]{zhang18b,yang23b}. For triple-aligned systems as suggested in this paper, an RM sign reversal is more difficult but not impossible. One possible configuration was presented as Case II left panel of Figure \ref{fig:three_figures}, with the right panel showing the extreme RM reversal observed in FRB 20190520B \citep{anna-thomas23}. One can see that for a triple aligned system with a large eccentricity, a companion with an inclined magnetic field would be viewed by the same observer to have opposite RM at different orbital phases. This is because the magnetar wind is likely pair-dominated with a negligible RM \citep{yang23b}. The dominant RM comes from the matter-dominated companion magnetized wind, whose line-of-sight magnetic field component has opposite signs near the perigee and apogee points. Within this picture, the orbital period is the period of the RM reversal cycle. Note that for some binaries with orbital angular momentum mis-aligned, the RM reversal is easier to achieve \citep{zhang18b,yang23b}. 
    \item RM flares: FRB 20220529 showed an abrupt increase and decrease of RM within weeks timescale, with the maximum amplitude more than an order of magnitude higher than the normal variation amplitude \citep{Li2025} (right panel of Case III of Figure \ref{fig:three_figures}). This behavior is best interpreted as the line of sight intersecting a corona mass ejection (CME)-like event similar to those ejected by the Sun \citep{Li2025}. Because an RM reversal was also observed, the orbit may be highly eccentric (or maybe it is not a triple-aligned system). See the left panel of Case III of Figure \ref{fig:three_figures}. 
\end{itemize}

\subsection{Propagation of FRB waves}\label{sec:propagation}

Even though the magnetospheric origin of FRB emission has been overwhelmingly supported by the observational data, the difficulties of propagating FRB waves within and slightly outside a magnetar magnetosphere have been discussed by various authors. \cite{beloborodov21} assumed that FRB waves are propagating through the closed field line region of a steady magnetar and showed that the waves cannot escape. \cite{qu22b} showed that this is not a problem for FRBs generated in the open field line region with a relativistically moving plasma (see also \citealt{lyutikov24}). \cite{sobacchi24} provided an even more stringent condition for the propagation of the FRB waves in a magnetar wind and suggested that FRBs have to be generated at a distance beyond $10^{12}$ cm from the engine. Their treatment also assumed that the FRB waves are propagating in the equatorial direction of a magnetar wind (i.e. the background $B$ field is perpendicular to the wave vector direction). It is worth pointing out that the outflow associated with FRB emission would likely significantly modify the magnetic field configuration along the propagation path, so the physical condition placed by \cite{sobacchi24} may not be relevant. In any case, in the aligned geometric configuration discussed in this paper, this difficulty is greatly eased because the FRB waves generated in the open field line region would continue to propagate in the open field line region without encountering a large $B_\perp$ before reaching a large distance. For a magnetar of a period of $\sim 2$ s, the light cylinder radius is $R_{\rm LC} \sim c P/2 \pi \sim 10^{10}$ cm. The FRB waves generated within the magnetosphere (e.g. at $10^8$ cm from the surface) can freely travel for $r = R_{\rm LC} / \sin\chi \sim 
(10^{12} \ {\rm cm})
\chi_{-2}^{-1}$ 
before hitting a significant $B_\perp$ component. At that radius, the damping effect discussed by \cite{sobacchi24} becomes negligibly small. So within our picture, prolific repeaters are likely (triple) aligned magnetars, and one-off events or inactive repeaters may originate from magnetars with other configurations in which most bursts (even if generated) could have been dissipated before escaping the magnetosphere and the near zone of the magnetar wind. 

\subsection{Companion-induced FRB triggers?}\label{sec:triggers}

Another important topic for active repeaters is their extremely high burst triggering rate. During active phases, these sources can emit bursts at a rate of a few hundred bursts per hour as observed with FAST \citep[e.g.][]{lid21,xuh22,zhangyk22,zhangyk23,Li2025,zhou25}. The record burst rate reaches $\sim 730 \ {\rm hr}^{-1}$ as detected by FAST \citep{zhangjs25}, which is more than one burst per 5 seconds. For an isolated magnetar, the mechanisms to trigger FRBs usually invoke star quakes or certain magnetic instabilities in the crust, with burst energy coming from strain, magnetic, rotational, and gravitational energies \citep[e.g.][]{wangwy24}.  However, for any burst associated with the release of strain or potential energy, the chance of triggering the next burst is dropped before strain or potential are built again. None of the known mechanisms can sustain continued energy release with such a high rate over a period of days to months. 

We speculate that the massive star companion of a magnetar may play a role of enhancing the burst triggers. For a massive star with mass loss rate of $\dot M \sim (10^{-3} {M_\odot ~ \rm yr^{-1}}) \dot M_{-3}$ and a wind speed of $v=0.01 \ c$ (or $\beta=0.01 \beta_{-2}$) located at a distance of $r = (10^{13} \ {\rm cm}) \ r_{13}$, the ram pressure at the location of the magnetar is $P_{\rm ram} = \rho v^2 = (\dot M v)/(4\pi r^2) \simeq (1.5 \times 10^4 \ {\rm erg \ cm^{-3}}) \ \dot M_{-3} \beta_{-2} r_{13}^{-2}$. For a typical magnetar with a spin period of $P$, the light cylinder radius is $R_{\rm lc} = c/\Omega = (4.8\times 10^9) P \ {\rm cm}$ (where $\Omega=2\pi/P$ is the angular velocity of the magnetar). The total kinetic energy deposit to the volumn of the magnetar is $E_{\rm kin} \sim P_{\rm ram} R_{\rm lc}^3 \sim 1.6 \times 10^{33} \ {\rm erg}$, which is much smaller than the energy of a typical FRB burst unless the individual bursts are extremely beamed with a beaming factor $f_b < 10^{-5}$. As a result, purely using the kinetic energy of the stellar wind to power FRBs \citep{zhang17} is not practical for the typical magnetar binaries discussed in this {\em Letter}. 

Nevertheless, the interaction between the companion wind and the magnetar magnetosphere may offer new possibilities to trigger FRBs. One possibility already raised by \cite{ioka20} was that the companion wind may send downward-streaming ``aurora'' particles into the magnetar magnetosphere, which may trigger two-stream instabilities when encountering outward moving particles and power coherent radio emission. Here we suggest another possibility: The companion wind may perturb the outer magnetosphere, sending Alfv\'en waves downward. The magnetic pressure at the light cylinder is $P_{\rm B}(R_{\rm lc})=B_{\rm lc}^2/8\pi = (B_s^2/8\pi)(\Omega R/c)^6 \simeq (3.4 \times 10^6 \ {\rm erg \ cm^{-3}}) B^2_{s,15} R_6^6 P^{-6}$ where the magnetar radius $R=10^6 \ {\rm cm} \ R_6$ and the its surface magnetic field is $B_s = 10^{15} \ {\rm G} \ B_{s,15}$. One can see $P_{\rm ram} / P_{\rm B} \sim 10^{-2}$, suggesting that the wind perturbation can trigger strong Alfv\'en waves. These waves may facilitate FRB emission in two possible ways. First, if the near-surface magnetic field configuration is metastable. The downward Alfv\'en waves may be able to trigger instability and induce reconnection, leading to release of a much larger amount of magnetic energy. Second, such an internal dissipation process will launch outgoing Alfv\'en waves. When the outgoing Alfv\'en waves collide with the downward-moving Alfv\'en waves perturbed by the companion wind, wave-wave interactions would lead to fast dissipation \citep[e.g.][]{lixy21} and possible particle acceleration. Coherent radio emission may be powered via coherent inverse Compton scattering of fast magnetosonic waves by relativistic particles \citep{zhang22,quzhang24} or by the free electron laser \citep{lyutikov21} mechanism.

If this speculation is correct, the very high burst rate of active repeaters in contrast of one-off FRBs can be attributed to two reasons: a nearly aligned geometry and an extra boost of the burst triggers by the companion. 

\subsection{Number density ratio between active repeaters and apparent non-repeaters}\label{sec:ratio}

Within our unified paradigm, the intrinsic source number density ratio between active repeaters and inactive repeaters (one-off or long-delay repeaters) can be estimated as
\begin{equation}
   \frac{n_{\rm a-rep}} {n_{\rm ina-rep}} \sim \frac{n_{\rm align} f_{\rm b,align}}{n_{\rm misalign} f_{\rm b,misalign}} \lesssim \frac{n_{\rm bi} f_{\rm b,align}}{n_{\rm iso} f_{\rm b,misalign}}, 
\end{equation}
where the number density ratio between magnetars with aligned and misaligned geometry, $n_{\rm align}/n_{\rm misalign}$, is somewhat smaller than the number density ratio between magnetars in binaries and in isolated systems, which for our fiducial model is $n_{\rm bi} /n_{\rm iso} \sim 4\% / 96\% \sim 0.04$. One can define $f_{\rm b} = \Delta\Omega / (4\pi)$ as the beaming correction factor of a magnetar, where $\Delta\Omega$ is the ``global'' solid angle of FRB emission, which is defined as the solid angle where all the bursts come from, not the beaming factor of the individual burst (see \cite{zhang23} for a definition). Assuming that FRB emission originates from a solid angle $\theta_c$ around the magnetic axis, an aligned rotator has a beaming factor $f_{\rm b,align} \sim 2(\pi \theta_c^2)/(4\pi) \sim \theta_c^2/2$. For a mis-aligned magnetar, on the other hand, the beaming factor may be estimated as $f_{\rm b,misalign} \sim 2\pi \theta_c \sin\chi / (4\pi) \sim \theta_c \sin\chi / 2$. One can see that $f_{\rm b,align}$ is generally smaller than $f_{\rm b,misalign}$ by a factor of $\theta_c \ll 1$, considering that $\sin\chi$ is of the order of unity. 

The detected number ratio between active repeaters and inactive repeaters, on the other hand, depends on many factors \citep[e.g.][]{ai21}, such as the energy function of FRBs, burst rate as a function of energy, the distributions of the energy function and the rate function, the detector's sensitivity and other detector-related parameters (zenith angle, operation duty cycle, etc.). A detailed Monte Carlo simulation is needed to address the full details, which is beyond the scope of this {\em Letter}. Here we estimate this ratio as
\begin{equation}
   \frac{N_{\rm a-rep}}{N_{\rm ina-rep}} \sim \frac{n_{\rm a-rep}} {n_{\rm ina-rep}} \frac{\left< \dot N_{\rm a-rep} \right>}{\left< \dot N_{\rm ina-rep} \right>} \lesssim \left( \frac{n_{\rm bi}}{n_{\rm iso}} \right) \theta_c \left< N_{\rm b,a-rep} \right>, 
\label{eq:Nratio}
\end{equation}
where $\left< \dot N_{\rm a-rep} \right>$ and $\left< \dot N_{\rm ina-rep} \right>$ are the average burst rates of active repeaters and inactive repeaters above the CHIME sensitivity threshold. This may be approximated as the ratio between the average detected bursts per FRB for active repeaters, $\left< N_{\rm b,a-rep} \right>$, and the average detected bursts per FRB for non-repeaters, which is 1. Observations from the CHIME FRB collaboration showed that $\left< N_{\rm b,a-rep} \right>$ is a few tens over the period of five years \citep{chime-new-repeaters}. Given a reasonable guess that $\theta_c$ is a few tenth, one can estimate that value of Eq.(\ref{eq:Nratio}) is of the order of $n_{\rm bi}/n_{\rm iso}$, which is a few percent, consistent with the observations \citep{chime-new-repeaters}. Future more observations and detailed modeling will verify this.

\section{Conclusions and discussion}\label{sec:conclusions}

Assuming that all FRBs in the universe are powered by magnetars, we propose that the majority of active repeaters are powered by magnetars in binary systems with a nearly aligned geometry between magnetic and spin axes ($\chi \sim 0$). By studying various magnetar formation channels, we suggest that such an aligned geometry can be easily achieved in a variety of magnetar formation channels involving Tayler-Spruit dynamo and tidal locking of the magnetar progenitor. However, the subsequent inclination angle $\chi$ evolution due to the spin-flip instability may remove such an alignment in isolated systems, as is the case for the known Galactic isolated magnetars. We argue that for some magnetars in binary systems with strong companion stellar winds, the aligned geometry can be sustained in a large fraction of binary magnetar systems, especially in those systems with the orbital angular momentum also aligned with the spin and magnetic vectors. We show that such triple aligned systems are likely the sources of most actively repeating FRBs. These systems can account for a variety of RM-evolution behaviors observed in active repeaters, including irregular and quasi-periodic RM evolutions, RM reversals, and RM flares. The geometry also facilitates the propagation of FRB waves generated within magnetar magnetospheres. We speculate that the existence of a companion may facilitate FRB triggers because of the interaction between the companion stellar wind and the magnetar magnetosphere. Such a scenario reasonably addresses some puzzling facts in FRB observations, including the lack of measured spin period, the diverse RM evolution behaviors, and the extremely high repetition rates of  active repeaters. 

The proposed physical picture does not rule out the existence of active repeaters in isolated magnetar systems or inactive repeaters in binary systems. As discussed in Section \ref{sec:alignment}, the majority of isolated magnetars are likely formed with an aligned geometry, but most would undergo $\chi$-evolution later due to spin-flip. The timescale for spin-flip to happen is very uncertain (which might be observationally constrained with a comprehensive analysis of Galactic magnetars in the future). It is possible that under favorable conditions, an isolated magnetar can maintain the nearly aligned configuration for an extended period, during which a large number of bursts are generated. FRB 20220912A had more than 1000 bursts detected by FAST in less than 9 hours \citep{zhangyk23} over a span of nearly two months. Its RM is essentially 0 and showed little variation during the 2- month timescale, suggesting a very clean environment. This source may be an isolated magnetar, or a magnetar with a companion that is not very magnetized and/or has no strong stellar winds. Conversely, magnetars in a binary system that have undergone spin flip (e.g. in B1 channel) would appear as an inactive repeater (one-off or low-repetition repeater) with a companion. If the red, faint source detected by JWST \citep{blanchard25} in association with the nearly CHIME-detected FRB 20250316A \citep{ng25} is indeed its counterpart, then the red source might be the companion star of this one-off FRB\footnote{Deep searches with the FAST telescope did not detect further bursts from the source \citep{qian25,liy25}.}. 

We only discussed the case that the companion is a stellar object. It is possible that the companion is an intermediate mass black hole (IMBH) \citep{zhang18b}, which can also eject a disk wind to provide the necessary magnetic field and plasma to account for the diverse RM evolution behaviors, and to interact with the magnetar magnetosphere to trigger FRBs. Several active repeaters have a putative orbital period of 100s of days \citep{rajwade20,pal25,liangyf25}. At such a large separation, an IMBH could launch a more powerful wind than a stellar companion, which may be more reasonable to interpret the observations \citep{wada21,yang23b}. However, the probability of having a magnetar engine near an IMBH and how it compares with that of aligned magnetars in binaries are not the subject of this study.

Finally, we only discussed the magnetar-for-all-FRBs scenario in this {\em Letter}. We have shown that it provides a reasonable description of the available FRB data. The competing scenario, that is, FRBs have multi-channel origins, is more difficult to test. Nonetheless, since magnetars are produced by a significant fraction of CCSNe, any formation channels that predict a much smaller rate density than CCSNe, e.g. compact star mergers, should be disfavored as the sole orgin of FRBs. Models involving accreting black holes to interpret repeating FRBs \citep[e.g.][]{katz17b,katz20,sridhar21} would be plausible if their source density is comparable to that of magnetars in binaries. Dedicated work is required to explore the suitability of various channels and their combinations to account for the FRB data.

\vspace{5mm}

The authors thank the anonymous referee for helpful suggestions, and Paz Beniamini, Dong Lai, Paul Lasky, Yuanhong Qu, Nanda Rea, Myles Sherman, Yihan Wang, and Yuan-Pei Yang for useful discussions or comments. 

\appendix

\section{The Magnetic Inclination Angle Evolution}
\label{sec:chi_evolution}
\setcounter{figure}{0}
\renewcommand{\thefigure}{A\arabic{figure}}

\begin{figure*}
\centering
\includegraphics[width=0.40\linewidth, trim = 0 0 0 0, clip]{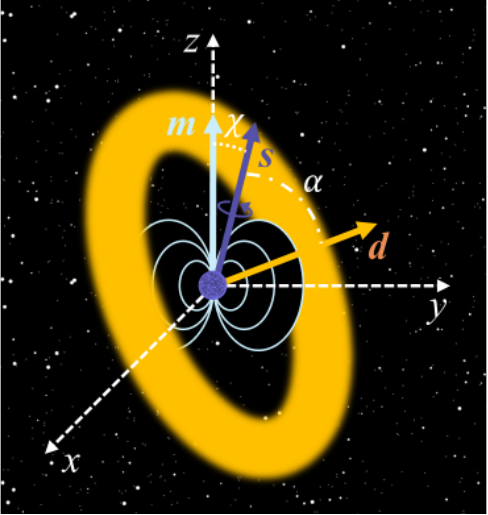}
\caption{A schematic illustration of the physical configuration surrounding an accreting magnetar. The central sphere represents the magnetar, surrounded by an orange accretion disk. Three unit vectors are shown to indicate key orientations: the $\bm{m}$ vector (light blue) represents the magnetic axis (z-axis), the $\bm{s}$ vector (dark blue) denotes the magnetar's spin axis, and the $\bm{d}$ vector (orange) is normal to the accretion disk plane (accretion disc axis).}
\label{fig:geometry}
\end{figure*}

The inclination angle $\chi$ between a magnetar's rotation axis and magnetic axis evolves over time due to internal viscous dissipation and external torques. 
In order to understand the influence of different torques on the inclination angle, we derive the evolution equation for the magnetic angle $\chi$ following \cite{biryukov21}.
Figure \ref{fig:geometry} shows the geometry: the $z$ axis is aligned with the magnetic axis $\bm{m}$, and the $x$ and $y$ axes are perpendicular to $z$ and form a right-handed coordinate system. The coordinate co-rotates with the star. The spin axis $\bm{s}$ lies in the $y-z$ plane, and the direction of the disk angular momentum $\bm{d}$ lies in an arbitrary direction. The unit vectors shown in Fig.\ref{fig:geometry} can be expressed as
\begin{equation}
    \bm{m} = 
    \begin{pmatrix}
    0 \\
    0 \\
    1
    \end{pmatrix}, ~~~~
    \bm{s} = 
    \begin{pmatrix}
    0 \\
    \sin\chi \\
    \cos\chi
    \end{pmatrix}, ~~~~
    \bm{d} = 
    \begin{pmatrix}
    \sin\alpha \sin\psi \\
    \sin\chi \cos\alpha - \cos\chi \sin\alpha \cos\psi \\
    \cos\chi \cos\alpha + \sin\chi \sin\alpha \cos\psi
    \end{pmatrix},
\end{equation}
where $\chi$ is the angle between the magnetic axis and the spin axis, $\alpha$ is the angle between the spin axis and the accretion disk axis, $\psi = -\int \Omega \mathrm{d}t$ is the rotational phase of the magnetar.

The rotational dynamics of the magnetar is governed by:
\begin{equation}
\label{eq: spin evolution}
I \frac{d\mathbf{\Omega}}{dt} + \frac{dI}{dt} \mathbf{\Omega} = \bm{N},
\end{equation}
where $I$ is the moment of inertia, $\mathbf{\Omega}$ is the angular velocity vector, $\bm{N} = \bm{N}_{\text{vsc}} + \bm{N}_{\text{ext}}$ is the total torque, which includes the internal viscous term and the external term.

Projecting onto $\bm{m}$, we get
\begin{equation}
I \frac{d}{dt} (\mathbf{\Omega} \cdot \bm{m}) = \bm{N} \cdot \bm{m} - \frac{dI}{dt} (\mathbf{\Omega} \cdot \bm{m}).
\end{equation}

Since $\mathbf{\Omega} \cdot \bm{m} = \Omega \cos\chi$, we have
\begin{equation}
\label{eq: chi}
I \left( \dot{\Omega} \cos\chi - \Omega \sin\chi \dot{\chi} \right) = \bm{N} \cdot \bm{m} - \frac{dI}{dt} \Omega \cos\chi. 
\end{equation}

Projecting Equation (\ref{eq: spin evolution}) onto axis $\bm{s}$, we obtain
\begin{equation}
I \frac{d}{dt} (\mathbf{\Omega} \cdot \bm{s}) = \bm{N} \cdot \bm{s} - \frac{dI}{dt} (\mathbf{\Omega} \cdot \bm{s}),
\end{equation}
which is simplified to 
\begin{equation}
I \dot{\Omega} = \bm{N} \cdot \bm{s} - \frac{dI}{dt} \Omega
\end{equation}
in view of $\mathbf{\Omega} \cdot \bm{s} = \Omega$.

Substituting it into Equation (\ref{eq: chi}), we can obtain an equation of the evolution of $\chi$ over time:
\begin{equation}
\label{eq: chi evolution}
I \Omega \sin\chi \dot{\chi} = \bm{N} \cdot \bm{s} \cos\chi - \bm{N} \cdot \bm{m}.
\end{equation}

\subsection{Isolated Magnetars}\label{sec:A-iso}

An isolated magnetar experiences an external torque caused by its magnetospheric current interaction with the neutron star itself, i.e. \citep{Spitkovsky2006,Philippov2014}
\begin{equation}
    \bm{N}_{\text{psr}} = K_{\text{psr}} \left\{ (1+\sin^2\chi) \bm{s} + ((\bm{m} \times \bm{s}) \times \bm{s}) \cos\chi \right\},
\end{equation}
which includes the spin-down torque (the first term) and a torque related to the evolutionary secular decrease of the obliquity (the second term). 

If we only consider the spin-down torque exerting on the magnetar ($\bm{N}=\bm{N}_{\text{psr}}$), one has
\begin{equation}
\bm{N}_{\text{psr}} \cdot \bm{s} = K_{\text{psr}} (1 + \sin^2\chi),
\end{equation}
and
\begin{equation}
\bm{N}_{\text{psr}} \cdot \bm{m} = K_{\text{psr}} \cos\chi.
\end{equation}
Substituting them into Equation (\ref{eq: chi evolution}), one gets
\begin{equation}
I \Omega \frac{d\chi}{dt} = K_{\text{psr}} \sin\chi \cos\chi.
\end{equation}
If the initial $\chi$ is not 0, one gets
\begin{equation}
    \frac{d}{dt} \ln \tan\chi = \frac{K_{\mathrm{psr}}}{I\Omega} = -\frac{\mu^2\Omega}{Ic^3},
\end{equation}
where $K_{\mathrm{psr}} = -\frac{\mu^2}{r_{\mathrm{LC}}^3}, r_{\mathrm{LC}} = c/\Omega$. One can see that the spin-down torque alone will continue to decrease the inclination angle $\chi$ toward zero, i.e. an $\bm{m}-\bm{s}$ aligned geometry.

However, the toroidal magnetic field of a magnetar is usually strong enough to deform the star into a prolate shape. When the rotation axis is misaligned with the magnetic axis, the initially hot star undergoes free precession, giving rise to internal viscous dissipation due to bulk viscosity. In the prolate case, the internal torque $\bm{N}_{\text{vsc}}$ caused by dissipation tends to increase the inclination angle $\chi$ toward $90 ^{\circ}$ \citep[i.e., evolving the magnetar towards an orthogonal configuration,][]{Lander2018}. 

Therefore, the spin-down torque $\bm{N}_{\text{psr}}$ competes against the internal viscous dissipation torque $\bm{N}_{\text{vsc}}$ during the early evolution of the magnetar. According to the detailed calculations of \cite{Lander2018}, an aligned geometry can be retained if the toroidal magnetic field is strong enough or the spin period is long enough; otherwise, the magnetar would evolve toward an orthogonal configuration (see, e.g.  Figure 14 of \citealt{Lander2018}). However, several uncertainties will further complicate the situation. For example, starquakes can change the inclination angle $\chi$, and a plasma-filled magnetosphere will introduce additional complexity, suggesting slower alignment timescales \citep{Philippov2014}.

If a magnetar does not reach an aligned or orthogonal configuration during the early stage, then the subsequent evolution of $\chi$ will be affected by other mechanisms operating in the later stages. This is particularly the case for magnetars in binary systems.

\subsection{Magnetars with a companion}\label{sec:A-bin}

If the magnetar has a companion, it may accrete matter during certain evolutionary stages when the companion fills its Roche lobe. In that case, both the accretion torque and the torque arising from the interaction between the magnetosphere and the accretion disk will act on the magnetar, the total external torque becomes $\bm{N} = \bm{N}_{\text{acc}} + \bm{N}_{\text{mag}} + \bm{N}_{\text{psr}}$.

Consider Equation (\ref{eq: chi evolution}) and averaging over the neutron star's spin period ($\langle x \rangle = \frac{1}{2\pi} \int^{2\pi}_{0} x \mathrm{d}\psi$):
\begin{equation}
\langle I \Omega \sin\chi \dot{\chi} \rangle = \langle \bm{N} \cdot \bm{s} \cos\chi \rangle - \langle \bm{N} \cdot \bm{m} \rangle.
\end{equation}

The accretion torque is $\bm{N}_{\text{acc}} = N_0 f(\cos\theta)\bm{d}$ \citep{biryukov21}, where $\theta$ denotes the angle between $\bm{m}$ and $\bm{d}$, i.e.
\begin{equation}
\cos\theta = \bm{m} \cdot \bm{d} = \cos\alpha \cos\chi + \sin\alpha \sin\chi \cos\psi,
\end{equation}
and the modulation function is
\begin{equation}
f(\cos\theta) = A(1 - \eta \cos^2\theta),
\end{equation}
where $\eta$ is a constant that describes the accretion torque modulation within the spin period, and $A$ is a normalization factor to make $\langle f(\cos\theta)\rangle = 1$, which reads
\begin{equation}
A = \left( \frac{1}{2\pi} \int^{2\pi}_{0} (1-\eta \cos^2\theta) \mathrm{d}\psi \right)^{-1}
= \left[ 1 - \frac{\eta}{2} (\sin^2\chi \sin^2\alpha + 2\cos^2\chi\cos^2\alpha) \right]^{-1}.
\end{equation}

The accretion torque contributes to
\begin{equation}
\langle \bm{N}_{\text{acc}} \cdot \bm{s} \rangle = N_0 \langle f(\cos\theta) (\bm{d} \cdot \bm{s}) \rangle
= N_0 \cos\alpha,
\end{equation}
and
\begin{align}
\langle \bm{N}_{\text{acc}} \cdot \bm{m} \rangle &= N_0 \langle f(\cos\theta) \cos\theta \rangle \\
&= N_0A\cos\alpha\cos\chi \left[1-\eta (\cos^2\alpha\cos^2\chi + \frac{3}{2}\sin^2\alpha\sin^2\chi)\right]\\
&= N_0 ( \cos\chi \cos\alpha - A \eta \sin^2\alpha \sin^2\chi \cos\alpha \cos\chi).
\end{align}

The magnetospheric torque $\bm{N}_{\text{mag}} = N_{\text{mag}} \bm{s}$ contributes to
\begin{equation}
\langle \bm{N}_{\text{mag}} \cdot \bm{s} \rangle = N_{\text{mag}},
\end{equation}
and 
\begin{equation}
\langle \bm{N}_{\text{mag}} \cdot \bm{m} \rangle = N_{\text{mag}} \cos\chi.
\end{equation}

Combining all terms together, one finally gets
\begin{equation}
\label{eq: MT_main}
I \Omega \left\langle \frac{d\chi}{dt} \right\rangle = \eta A N_0 \sin^2\alpha \cos\alpha \sin\chi \cos\chi + K_{\text{psr}} \sin\chi \cos\chi.
\end{equation}

In wind-fed accretion systems (the companion star has a strong stellar wind), the angular momentum direction of the accreted materials fluctuates stochastically due to inhomogeneities in the donor star's wind. This fundamentally alters the magnetic angle evolution \citep{biryukov21}.

The accretion torque maintains its form:
\begin{equation}
\bm{N}_{\text{acc}} = N_0 f(\cos\theta)\bm{d}(t)
\end{equation}
but now with a stochastic $\bm{d}(t)$.

Consider the evolution in a long timescale. If we neglect pulsar losses ($K_{\text{psr}} \approx 0$), Equation (\ref{eq: MT_main}) becomes
\begin{equation}
\label{eq: wind}
\frac{d}{dt} \ln \tan\chi = \eta A \frac{N_0}{I\Omega} \sin^2\alpha(t) \cos\alpha(t)
\end{equation}

We can think of the evolution of $\ln \tan\chi$ as a random walk process caused by a white noise on the right-hand side of Equation (\ref{eq: wind}). The dispersion of the quantity reads
\begin{equation}
\label{eq: dispersion}
\langle (\ln \tan\chi)^2 \rangle \simeq \eta^2 A^2 \left(\frac{N_0}{I\Omega}\right)^2 \langle (\sin^2\alpha \cos\alpha)^2 \rangle \tau_{\alpha}t.
\end{equation}

\subsubsection{Random alignment}
We first consider that $\alpha$ is uniformly distributed over the sphere, corresponding to the case where the spin and orbital angular momentum vector angle has a random distribution. The probability density function for $\alpha$ is
\begin{equation}
P(\alpha) = \frac{\sin\alpha}{2}, \alpha \in [0,\pi],
\end{equation}
so, the average is computed as
\begin{align}
\langle (\sin^2\alpha \cos\alpha)^2 \rangle 
&= \int^{\pi}_{0} \sin^4\alpha \cos^2\alpha P(\alpha)\mathrm{d}\alpha
= \frac{1}{2} \int^{\pi}_{0} \sin^5\alpha \cos^2\alpha \mathrm{d}\alpha 
= \frac{8}{105}.
\end{align}

Equation (\ref{eq: dispersion}) becomes:
\begin{equation}
\langle (\ln \tan\chi)^2 \rangle \simeq \frac{8}{105} \eta^2 A^2 \left(\frac{N_0}{I\Omega}\right)^2 \tau_{\alpha}t.
\end{equation}
One can see that the magnetic angle is involved in a random walk process that is linear in $\ln(\tan\chi)$, and thus tends to form a uniform distribution in this quantity. A uniform distribution in $\ln(\tan\chi)$ corresponds to
\begin{equation}
    \frac{\mathrm{d}N}{\mathrm{d}\chi} \propto \frac{1}{|\sin\chi\cos\chi|},
\end{equation}
which peaks at $0 ^{\circ}$ or $90 ^{\circ}$. 

\subsubsection{Triple alignment}

Consider a magnetar whose spin axis is nearly parallel to the orbital angular momentum, and suppose the magnetic axis is initially only misaligned by a small angle (i.e. a triple-aligned system). Since the binary separation is much larger than the radius of the magnetar, the direction of accretion torque is considered to be nearly parallel (or antiparallel) to the direction of orbital angular momentum, i.e. $\alpha \sim 0$ or $\pi$.

Following Equation (\ref{eq: wind}), let us define
\begin{equation}
    f(t) = \eta A \frac{N_0}{I\Omega} \sin^2\alpha(t) \cos\alpha(t).
\end{equation}
When $\alpha \simeq 0$, one has
\begin{equation}
    f(t) \simeq \eta A \frac{N_0}{I\Omega} \alpha^2;
\end{equation}
and when $\alpha \simeq \pi$, one has
\begin{equation}
    f(t) \simeq -\eta A \frac{N_0}{I\Omega} \epsilon^2,
\end{equation}
where $\epsilon=\pi-\alpha$.

Since the initial $\chi \sim 0$, we have $\tan \chi \simeq \chi$. Equation (\ref{eq: wind}) becomes
\begin{equation}
    \chi(t) = \chi(0) e^{\int f(t) dt}.
\end{equation}
The evolution of $\chi$ is thus exponentially sensitive to the time-integrated fluctuations.

If $\chi$ increases, then the evolution of $\chi$ would become
\begin{equation}
    \tan \chi(t) = \tan \chi(0) e^{\int f(t) dt},
\end{equation}
indicating that the growth rate of $\chi$ becomes slower than a purely exponential relationship.

When the aligned configuration lies very close to the starting point in the angle space, the magnetar is more likely to reach the aligned configuration than to be driven to the orthogonal configuration.


\begin{thebibliography}{}
\expandafter\ifx\csname natexlab\endcsname\relax\def\natexlab#1{#1}\fi
\providecommand{\url}[1]{\href{#1}{#1}}
\providecommand{\dodoi}[1]{doi:~\href{http://doi.org/#1}{\nolinkurl{#1}}}
\providecommand{\doeprint}[1]{\href{http://ascl.net/#1}{\nolinkurl{http://ascl.net/#1}}}
\providecommand{\doarXiv}[1]{\href{https://arxiv.org/abs/#1}{\nolinkurl{https://arxiv.org/abs/#1}}}

\bibitem[{{Ablimit}(2022)}]{Ablimit2022a}
{Ablimit}, I. 2022, \mnras, 509, 6061, \dodoi{10.1093/mnras/stab3060}

\bibitem[{{Ablimit} {et~al.}(2022){Ablimit}, {Podsiadlowski}, {Hirai}, \&
  {Wicker}}]{Ablimit2022b}
{Ablimit}, I., {Podsiadlowski}, P., {Hirai}, R., \& {Wicker}, J. 2022, \mnras,
  513, 4802, \dodoi{10.1093/mnras/stac631}

\bibitem[{{Ai} {et~al.}(2021){Ai}, {Gao}, \& {Zhang}}]{ai21}
{Ai}, S., {Gao}, H., \& {Zhang}, B. 2021, \apjl, 906, L5,
  \dodoi{10.3847/2041-8213/abcec9}

\bibitem[{{Anna-Thomas} {et~al.}(2023{\natexlab{a}}){Anna-Thomas}, {Connor},
  {Dai}, {Feng}, {Burke-Spolaor}, {Beniamini}, {Yang}, {Zhang}, {Aggarwal},
  {Law}, {Li}, {Niu}, {Chatterjee}, {Cruces}, {Duan}, {Filipovic}, {Hobbs},
  {Lynch}, {Miao}, {Niu}, {Ocker}, {Tsai}, {Wang}, {Xue}, {Yao}, {Yu}, {Zhang},
  {Zhang}, {Zhu}, \& {Zhu}}]{anna-thomas23}
{Anna-Thomas}, R., {Connor}, L., {Dai}, S., {et~al.} 2023{\natexlab{a}},
  Science, 380, 599, \dodoi{10.1126/science.abo6526}

\bibitem[{{Anna-Thomas} {et~al.}(2023{\natexlab{b}}){Anna-Thomas}, {Connor},
  {Dai}, {Feng}, {Burke-Spolaor}, {Beniamini}, {Yang}, {Zhang}, {Aggarwal},
  {Law}, {Li}, {Niu}, {Chatterjee}, {Cruces}, {Duan}, {Filipovic}, {Hobbs},
  {Lynch}, {Miao}, {Niu}, {Ocker}, {Tsai}, {Wang}, {Xue}, {Yao}, {Yu}, {Zhang},
  {Zhang}, {Zhu}, \& {Zhu}}]{Anna2023}
---. 2023{\natexlab{b}}, Science, 380, 599, \dodoi{10.1126/science.abo6526}

\bibitem[{{Barr{\`e}re} {et~al.}(2022){Barr{\`e}re}, {Guilet}, {Reboul-Salze},
  {Raynaud}, \& {Janka}}]{Barrere2022}
{Barr{\`e}re}, P., {Guilet}, J., {Reboul-Salze}, A., {Raynaud}, R., \& {Janka},
  H.~T. 2022, \aap, 668, A79, \dodoi{10.1051/0004-6361/202244172}

\bibitem[{{Beloborodov}(2021)}]{beloborodov21}
{Beloborodov}, A.~M. 2021, \apjl, 922, L7, \dodoi{10.3847/2041-8213/ac2fa0}

\bibitem[{{Beniamini} {et~al.}(2019){Beniamini}, {Hotokezaka}, {van der Horst},
  \& {Kouveliotou}}]{Beniamini2019}
{Beniamini}, P., {Hotokezaka}, K., {van der Horst}, A., \& {Kouveliotou}, C.
  2019, \mnras, 487, 1426, \dodoi{10.1093/mnras/stz1391}

\bibitem[{{Beniamini} \& {Kumar}(2025)}]{beniamini25}
{Beniamini}, P., \& {Kumar}, P. 2025, \apj, 982, 45,
  \dodoi{10.3847/1538-4357/adb8e6}

\bibitem[{{Beniamini} \& {Piran}(2024)}]{beniamini24}
{Beniamini}, P., \& {Piran}, T. 2024, \apj, 966, 17,
  \dodoi{10.3847/1538-4357/ad32cd}

\bibitem[{{Biryukov} \& {Abolmasov}(2021)}]{biryukov21}
{Biryukov}, A., \& {Abolmasov}, P. 2021, \mnras, 505, 1775,
  \dodoi{10.1093/mnras/stab1378}

\bibitem[{{Blanchard} {et~al.}(2025){Blanchard}, {Berger}, {Hiramatsu},
  {Golay}, {Kumar}, {Metzger}, {Sridhar}, \& Kohki}]{blanchard25}
{Blanchard}, P.~K., {Berger}, E., {Hiramatsu}, D., {et~al.} 2025, The
  Astronomer's Telegram, 17195, 1

\bibitem[{{Bochenek} {et~al.}(2020){Bochenek}, {Ravi}, {Belov}, {Hallinan},
  {Kocz}, {Kulkarni}, \& {McKenna}}]{STARE2-SGR}
{Bochenek}, C.~D., {Ravi}, V., {Belov}, K.~V., {et~al.} 2020, \nat, 587, 59,
  \dodoi{10.1038/s41586-020-2872-x}

\bibitem[{{Camilo} {et~al.}(2007){Camilo}, {Cognard}, {Ransom}, {Halpern},
  {Reynolds}, {Zimmerman}, {Gotthelf}, {Helfand}, {Demorest}, {Theureau}, \&
  {Backer}}]{camilo07}
{Camilo}, F., {Cognard}, I., {Ransom}, S.~M., {et~al.} 2007, \apj, 663, 497,
  \dodoi{10.1086/518226}

\bibitem[{{CHIME/FRB Collaboration} {et~al.}(2020){CHIME/FRB Collaboration},
  {Andersen}, {Bandura}, {Bhardwaj}, {Bij}, {Boyce}, {Boyle}, {Brar},
  {Cassanelli}, {Chawla}, {Chen}, {Cliche}, {Cook}, {Cubranic}, {Curtin},
  {Denman}, {Dobbs}, {Dong}, {Fandino}, {Fonseca}, {Gaensler}, {Giri}, {Good},
  {Halpern}, {Hill}, {Hinshaw}, {H{\"o}fer}, {Josephy}, {Kania}, {Kaspi},
  {Landecker}, {Leung}, {Li}, {Lin}, {Masui}, {McKinven}, {Mena-Parra},
  {Merryfield}, {Meyers}, {Michilli}, {Milutinovic}, {Mirhosseini},
  {M{\"u}nchmeyer}, {Naidu}, {Newburgh}, {Ng}, {Patel}, {Pen},
  {Pinsonneault-Marotte}, {Pleunis}, {Quine}, {Rafiei-Ravandi}, {Rahman},
  {Ransom}, {Renard}, {Sanghavi}, {Scholz}, {Shaw}, {Shin}, {Siegel}, {Singh},
  {Smegal}, {Smith}, {Stairs}, {Tan}, {Tendulkar}, {Tretyakov}, {Vanderlinde},
  {Wang}, {Wulf}, \& {Zwaniga}}]{CHIME-SGR}
{CHIME/FRB Collaboration}, {Andersen}, B.~C., {Bandura}, K.~M., {et~al.} 2020,
  \nat, 587, 54, \dodoi{10.1038/s41586-020-2863-y}

\bibitem[{{CHIME/FRB Collaboration} {et~al.}(2021){CHIME/FRB Collaboration},
  {Amiri}, {Andersen}, {Bandura}, {Berger}, {Bhardwaj}, {Boyce}, {Boyle},
  {Brar}, {Breitman}, {Cassanelli}, {Chawla}, {Chen}, {Cliche}, {Cook},
  {Cubranic}, {Curtin}, {Deng}, {Dobbs}, {Dong}, {Eadie}, {Fandino}, {Fonseca},
  {Gaensler}, {Giri}, {Good}, {Halpern}, {Hill}, {Hinshaw}, {Josephy},
  {Kaczmarek}, {Kader}, {Kania}, {Kaspi}, {Landecker}, {Lang}, {Leung}, {Li},
  {Lin}, {Masui}, {McKinven}, {Mena-Parra}, {Merryfield}, {Meyers}, {Michilli},
  {Milutinovic}, {Mirhosseini}, {M{\"u}nchmeyer}, {Naidu}, {Newburgh}, {Ng},
  {Patel}, {Pen}, {Petroff}, {Pinsonneault-Marotte}, {Pleunis},
  {Rafiei-Ravandi}, {Rahman}, {Ransom}, {Renard}, {Sanghavi}, {Scholz}, {Shaw},
  {Shin}, {Siegel}, {Sikora}, {Singh}, {Smith}, {Stairs}, {Tan}, {Tendulkar},
  {Vanderlinde}, {Wang}, {Wulf}, \& {Zwaniga}}]{chime-1st-catalog}
{CHIME/FRB Collaboration}, {Amiri}, M., {Andersen}, B.~C., {et~al.} 2021,
  \apjs, 257, 59, \dodoi{10.3847/1538-4365/ac33ab}

\bibitem[{{Chime/Frb Collaboration} {et~al.}(2023){Chime/Frb Collaboration},
  {Andersen}, {Bandura}, {Bhardwaj}, {Boyle}, {Brar}, {Cassanelli},
  {Chatterjee}, {Chawla}, {Cook}, {Curtin}, {Dobbs}, {Dong}, {Faber},
  {Fandino}, {Fonseca}, {Gaensler}, {Giri}, {Herrera-Martin}, {Hill}, {Ibik},
  {Josephy}, {Kaczmarek}, {Kader}, {Kaspi}, {Landecker}, {Lanman}, {Lazda},
  {Leung}, {Lin}, {Masui}, {McKinven}, {Mena-Parra}, {Meyers}, {Michilli},
  {Ng}, {Pandhi}, {Pearlman}, {Pen}, {Petroff}, {Pleunis}, {Rafiei-Ravandi},
  {Rahman}, {Ransom}, {Renard}, {Sand}, {Sanghavi}, {Scholz}, {Shah}, {Shin},
  {Siegel}, {Smith}, {Stairs}, {Su}, {Tendulkar}, {Vanderlinde}, {Wang},
  {Wulf}, \& {Zwaniga}}]{chime-new-repeaters}
{Chime/Frb Collaboration}, {Andersen}, B.~C., {Bandura}, K., {et~al.} 2023,
  \apj, 947, 83, \dodoi{10.3847/1538-4357/acc6c1}

\bibitem[{{Chime/Frb Collaboration} {et~al.}(2022){Chime/Frb Collaboration},
  {Bandura}, {Bhardwaj}, {Boyle}, {Brar}, {Breitman}, {Cassanelli},
  {Chatterjee}, {Chawla}, {Cliche}, {Cubranic}, {Curtin}, {Deng}, {Dobbs},
  {Dong}, {Fonseca}, {Gaensler}, {Giri}, {Good}, {Hill}, {Josephy},
  {Kaczmarek}, {Kader}, {Kania}, {Kaspi}, {Leung}, {Li}, {Lin}, {Masui},
  {McKinven}, {Mena-Parra}, {Merryfield}, {Meyers}, {Michilli}, {Naidu},
  {Newburgh}, {Ng}, {Ordog}, {Patel}, {Pearlman}, {Pen}, {Petroff}, {Pleunis},
  {Rafiei-Ravandi}, {Rahman}, {Ransom}, {Renard}, {Sanghavi}, {Scholz}, {Shaw},
  {Shin}, {Siegel}, {Singh}, {Smith}, {Stairs}, {Tan}, {Tendulkar},
  {Vanderlinde}, {Wiebe}, {Wulf}, \& {Zwaniga}}]{chime-period}
{Chime/Frb Collaboration}, Andersen, B.~C., {Bandura}, K., {Bhardwaj}, M.,
  {et~al.} 2022, \nat, 607, 256, \dodoi{10.1038/s41586-022-04841-8}

\bibitem[{{Chrimes} {et~al.}(2022){Chrimes}, {Levan}, {Fruchter}, {Groot},
  {Jonker}, {Kouveliotou}, {Lyman}, {Stanway}, {Tanvir}, \&
  {Wiersema}}]{Chrimes2022}
{Chrimes}, A.~A., {Levan}, A.~J., {Fruchter}, A.~S., {et~al.} 2022, \mnras,
  513, 3550, \dodoi{10.1093/mnras/stac1090}

\bibitem[{{Cutler}(2002)}]{cutler02}
{Cutler}, C. 2002, \prd, 66, 084025, \dodoi{10.1103/PhysRevD.66.084025}

\bibitem[{{Dai} {et~al.}(2006){Dai}, {Wang}, {Wu}, \& {Zhang}}]{dai06}
{Dai}, Z.~G., {Wang}, X.~Y., {Wu}, X.~F., \& {Zhang}, B. 2006, Science, 311,
  1127, \dodoi{10.1126/science.1123606}

\bibitem[{{Doroshenko} {et~al.}(2021){Doroshenko}, {Santangelo}, {Tsygankov},
  \& {Ji}}]{Doroshenko21}
{Doroshenko}, V., {Santangelo}, A., {Tsygankov}, S.~S., \& {Ji}, L. 2021, \aap,
  647, A165, \dodoi{10.1051/0004-6361/202039785}

\bibitem[{{Duncan} \& {Thompson}(1992)}]{duncan92}
{Duncan}, R.~C., \& {Thompson}, C. 1992, \apjl, 392, L9, \dodoi{10.1086/186413}

\bibitem[{{Ferrario} \& {Wickramasinghe}(2006)}]{ferrario06}
{Ferrario}, L., \& {Wickramasinghe}, D. 2006, \mnras, 367, 1323,
  \dodoi{10.1111/j.1365-2966.2006.10058.x}

\bibitem[{{Frost} {et~al.}(2024){Frost}, {Sana}, {Mahy}, {Wade}, {Barron}, {Le
  Bouquin}, {M{\'e}rand}, {Schneider}, {Shenar}, {Barb{\'a}}, {Bowman},
  {Fabry}, {Farhang}, {Marchant}, {Morrell}, \& {Smoker}}]{Frost2024}
{Frost}, A.~J., {Sana}, H., {Mahy}, L., {et~al.} 2024, Science, 384, 214,
  \dodoi{10.1126/science.adg7700}

\bibitem[{{Gao} {et~al.}(2013){Gao}, {Ding}, {Wu}, {Zhang}, \& {Dai}}]{gao13}
{Gao}, H., {Ding}, X., {Wu}, X.-F., {Zhang}, B., \& {Dai}, Z.-G. 2013, \apj,
  771, 86, \dodoi{10.1088/0004-637X/771/2/86}

\bibitem[{{Gao} {et~al.}(2016){Gao}, {Zhang}, \& {L{\"u}}}]{gao16}
{Gao}, H., {Zhang}, B., \& {L{\"u}}, H.-J. 2016, \prd, 93, 044065,
  \dodoi{10.1103/PhysRevD.93.044065}

\bibitem[{{Gao} \& {Fan}(2006)}]{gaofan06}
{Gao}, W.-H., \& {Fan}, Y.-Z. 2006, \cjaa, 6, 513,
  \dodoi{10.1088/1009-9271/6/5/01}

\bibitem[{{Hu} {et~al.}(2023){Hu}, {Zhu}, {Qin}, {Shao}, {Zhang}, {Yu},
  {Liang}, {Liu}, {Wang}, {Shu}, \& {Liu}}]{Hu2023}
{Hu}, R.-C., {Zhu}, J.-P., {Qin}, Y., {et~al.} 2023, arXiv e-prints,
  arXiv:2301.06402, \dodoi{10.48550/arXiv.2301.06402}

\bibitem[{{Ioka} \& {Zhang}(2020)}]{ioka20}
{Ioka}, K., \& {Zhang}, B. 2020, \apjl, 893, L26,
  \dodoi{10.3847/2041-8213/ab83fb}

\bibitem[{{Janka}(2012)}]{janka12}
{Janka}, H.-T. 2012, Annual Review of Nuclear and Particle Science, 62, 407,
  \dodoi{10.1146/annurev-nucl-102711-094901}

\bibitem[{{Janka} {et~al.}(2022){Janka}, {Wongwathanarat}, \&
  {Kramer}}]{Janka2022}
{Janka}, H.-T., {Wongwathanarat}, A., \& {Kramer}, M. 2022, \apj, 926, 9,
  \dodoi{10.3847/1538-4357/ac403c}

\bibitem[{{Jiang} {et~al.}(2022){Jiang}, {Wang}, {Xu}, {Xu}, {Zhang}, {Wang},
  {Zhou}, {Zhang}, {Niu}, {Lee}, {Zhang}, {Han}, {Li}, {Zhu}, {Dai}, {Feng},
  {Jing}, {Li}, {Luo}, {Miao}, {Niu}, {Tsai}, {Wang}, {Wang}, {Xu}, {Yang},
  {Yang}, {Yao}, \& {Yuan}}]{jiangjc22}
{Jiang}, J.-C., {Wang}, W.-Y., {Xu}, H., {et~al.} 2022, Research in Astronomy
  and Astrophysics, 22, 124003, \dodoi{10.1088/1674-4527/ac98f6}

\bibitem[{{Jiang} {et~al.}(2024){Jiang}, {Xu}, {Niu}, {Lee}, {Zhu}, {Zhang},
  {Qu}, {Xu}, {Zhou}, {Cao}, {Wang}, {Wang}, {Cao}, {Zhang}, {Zhang}, {Gan},
  {Han}, {Hao}, {Huang}, {Jiang}, {Li}, {Li}, {Li}, {Li}, {Luo}, {Men}, {Qian},
  {Sun}, {Wang}, {Xu}, {Xu}, {Yang}, {Yao}, {Yue}, {Yu}, {Yuan}, \&
  {Zhu}}]{jiangjc24}
{Jiang}, J.~C., {Xu}, J.~W., {Niu}, J.~R., {et~al.} 2024, National Science
  Review, 12, nwae293, \dodoi{10.1093/nsr/nwae293}

\bibitem[{{Kaspi} \& {Beloborodov}(2017)}]{kaspi17}
{Kaspi}, V.~M., \& {Beloborodov}, A.~M. 2017, \araa, 55, 261,
  \dodoi{10.1146/annurev-astro-081915-023329}

\bibitem[{{Katz}(2017)}]{katz17b}
{Katz}, J.~I. 2017, \mnras, 471, L92, \dodoi{10.1093/mnrasl/slx113}

\bibitem[{{Katz}(2020)}]{katz20}
---. 2020, \mnras, 494, L64, \dodoi{10.1093/mnrasl/slaa038}

\bibitem[{{Kirsten} {et~al.}(2021){Kirsten}, {Snelders}, {Jenkins}, {Nimmo},
  {van den Eijnden}, {Hessels}, {Gawro{\'n}ski}, \& {Yang}}]{kirsten21}
{Kirsten}, F., {Snelders}, M.~P., {Jenkins}, M., {et~al.} 2021, Nature
  Astronomy, 5, 414, \dodoi{10.1038/s41550-020-01246-3}

\bibitem[{{Kumar} \& {Bo{\v{s}}njak}(2020)}]{kumar20a}
{Kumar}, P., \& {Bo{\v{s}}njak}, {\v{Z}}. 2020, \mnras, 494, 2385,
  \dodoi{10.1093/mnras/staa774}

\bibitem[{{Kumar} {et~al.}(2017){Kumar}, {Lu}, \& {Bhattacharya}}]{kumar17}
{Kumar}, P., {Lu}, W., \& {Bhattacharya}, M. 2017, \mnras, 468, 2726,
  \dodoi{10.1093/mnras/stx665}

\bibitem[{{Lai} {et~al.}(1993){Lai}, {Rasio}, \& {Shapiro}}]{lai93}
{Lai}, D., {Rasio}, F.~A., \& {Shapiro}, S.~L. 1993, \apjs, 88, 205,
  \dodoi{10.1086/191822}

\bibitem[{{Lai} {et~al.}(1994){Lai}, {Rasio}, \& {Shapiro}}]{lai94}
---. 1994, \apj, 420, 811, \dodoi{10.1086/173606}

\bibitem[{{Lander} \& {Jones}(2018)}]{Lander2018}
{Lander}, S.~K., \& {Jones}, D.~I. 2018, \mnras, 481, 4169,
  \dodoi{10.1093/mnras/sty2553}

\bibitem[{{Lasky} \& {Glampedakis}(2016)}]{lasky16}
{Lasky}, P.~D., \& {Glampedakis}, K. 2016, \mnras, 458, 1660,
  \dodoi{10.1093/mnras/stw435}

\bibitem[{{Li} {et~al.}(2021{\natexlab{a}}){Li}, {Lin}, {Xiong}, {Ge}, {Li},
  {Li}, {Lu}, {Zhang}, {Tuo}, {Nang}, {Zhang}, {Xiao}, {Chen}, {Song}, {Xu},
  {Liu}, {Jia}, {Cao}, {Qu}, {Zhang}, {Gu}, {Liao}, {Zhao}, {Tan}, {Nie},
  {Zhao}, {Zheng}, {Zheng}, {Luo}, {Cai}, {Li}, {Xue}, {Bu}, {Chang}, {Chen},
  {Chen}, {Chen}, {Chen}, {Chen}, {Cui}, {Cui}, {Deng}, {Dong}, {Du}, {Fu},
  {Gao}, {Gao}, {Gao}, {Gu}, {Guan}, {Guo}, {Han}, {Huang}, {Huo}, {Jiang},
  {Jiang}, {Jin}, {Jin}, {Kong}, {Li}, {Li}, {Li}, {Li}, {Li}, {Li}, {Li},
  {Liang}, {Liu}, {Liu}, {Liu}, {Liu}, {Liu}, {Lu}, {Lu}, {Luo}, {Ma}, {Meng},
  {Ou}, {Sai}, {Shang}, {Song}, {Sun}, {Tao}, {Wang}, {Wang}, {Wang}, {Wang},
  {Wang}, {Wen}, {Wu}, {Wu}, {Wu}, {Xiao}, {Xu}, {Yang}, {Yang}, {Yang},
  {Yang}, {Yi}, {Yin}, {You}, {Zhang}, {Zhang}, {Zhang}, {Zhang}, {Zhang},
  {Zhang}, {Zhang}, {Zhang}, {Zhang}, {Zhou}, {Zhou}, {Zhu}, {Zhu}, \&
  {Zhuang}}]{HXMT-SGR}
{Li}, C.~K., {Lin}, L., {Xiong}, S.~L., {et~al.} 2021{\natexlab{a}}, Nature
  Astronomy, 5, 378, \dodoi{10.1038/s41550-021-01302-6}

\bibitem[{{Li} {et~al.}(2021{\natexlab{b}}){Li}, {Wang}, {Zhu}, {Zhang},
  {Zhang}, {Duan}, {Zhang}, {Feng}, {Tang}, {Chatterjee}, {Cordes}, {Cruces},
  {Dai}, {Gajjar}, {Hobbs}, {Jin}, {Kramer}, {Lorimer}, {Miao}, {Niu}, {Niu},
  {Pan}, {Qian}, {Spitler}, {Werthimer}, {Zhang}, {Wang}, {Xie}, {Yue},
  {Zhang}, {Zhi}, \& {Zhu}}]{lid21}
{Li}, D., {Wang}, P., {Zhu}, W.~W., {et~al.} 2021{\natexlab{b}}, \nat, 598,
  267, \dodoi{10.1038/s41586-021-03878-5}

\bibitem[{{Li} {et~al.}(2021{\natexlab{c}}){Li}, {Beloborodov}, \&
  {Sironi}}]{lixy21}
{Li}, X., {Beloborodov}, A.~M., \& {Sironi}, L. 2021{\natexlab{c}}, \apj, 915,
  101, \dodoi{10.3847/1538-4357/abfe5f}

\bibitem[{{Li} {et~al.}(2025{\natexlab{a}}){Li}, {Zhang}, {Yang}, {Tsai},
  {Yang}, {Law}, {Anna-Thomas}, {Chen}, {Lee}, {Tang}, {Xiao}, {Xu}, {Yang},
  {Chen}, {Feng}, {Li}, {Mckinven}, {Niu}, {Shin}, {Wang}, {Zhang}, {Zhang},
  {Zhou}, {Zhu}, {Dai}, {Chang}, {Geng}, {Han}, {Hu}, {Li}, {Luo}, {Niu},
  {Shi}, {Sun}, {Wu}, {Zhu}, {Jiang}, \& {Zhang}}]{Li2025}
{Li}, Y., {Zhang}, S.~B., {Yang}, Y.~P., {et~al.} 2025{\natexlab{a}}, arXiv
  e-prints, arXiv:2503.04727, \dodoi{10.48550/arXiv.2503.04727}

\bibitem[{{Li} {et~al.}(2025{\natexlab{b}}){Li}, {Sun}, {Qian}, {Li}, {Hua},
  {Xin}, {Li}, {Wang}, {Niu}, {Sun}, {Yao}, {Geng}, {Jin}, {Rea}, {Liu}, {Pan},
  {An}, {Burwitz}, {Cai}, {Cao}, {Chen}, {Cheng}, {Cui}, {Feng}, {Friedrich},
  {Han}, {Hu}, {Hu}, {Huang}, {Jia}, {Jiang}, {Li}, {Li}, {Liang}, {Liang},
  {Liu}, {Liu}, {Liu}, {Meidinger}, {Pan}, {Rau}, {Shu}, {Sun}, {Tao}, {Tang},
  {Wan}, {Wang}, {Wang}, {Wang}, {Xu}, {Xue}, {Yang}, {Yao}, {Yao}, {Zhao},
  {Zhao}, {Zhang}, {Zhang}, {Zhang}, {Zhang}, {Zhang}, {Zhang}, {Zhang},
  {Zhang}, {Zhang}, {Zheng}, {Zhu}, {Zuo}, {Sun}, {Wei}, {Zhu}, {Jiang},
  {Yuan}, {Wu}, \& {Zhang}}]{liy25}
{Li}, Y., {Sun}, H., {Qian}, L., {et~al.} 2025{\natexlab{b}}, arXiv e-prints,
  arXiv:2508.13999, \dodoi{10.48550/arXiv.2508.13999}

\bibitem[{{Liang} {et~al.}(2025){Liang}, {Li}, {Tang}, {Yang}, {Zhang}, {Yang},
  {Wang}, {Wang}, {Xiao}, {Zhao}, {Wei}, {Geng}, {Niu}, {Zhang}, {Chen},
  {Fang}, {Wu}, {Dai}, {Zhu}, {Jiang}, \& {Zhang}}]{liangyf25}
{Liang}, Y.-F., {Li}, Y., {Tang}, Z.-F., {et~al.} 2025, arXiv e-prints,
  arXiv:2505.10463, \dodoi{10.48550/arXiv.2505.10463}

\bibitem[{{Lin} {et~al.}(2020){Lin}, {Zhang}, {Wang}, {Gao}, {Guan}, {Han},
  {Jiang}, {Jiang}, {Lee}, {Li}, {Men}, {Miao}, {Niu}, {Niu}, {Sun}, {Wang},
  {Wang}, {Xu}, {Xu}, {Xu}, {Yang}, {Yang}, {Yu}, {Zhang}, {Zhang}, {Zhou},
  {Zhu}, {Castro-Tirado}, {Dai}, {Ge}, {Hu}, {Li}, {Li}, {Li}, {Liang}, {Jia},
  {Querel}, {Shao}, {Wang}, {Wang}, {Wu}, {Xiong}, {Xu}, {Yang}, {Zhang},
  {Zhang}, {Zheng}, \& {Zou}}]{lin20}
{Lin}, L., {Zhang}, C.~F., {Wang}, P., {et~al.} 2020, \nat, 587, 63,
  \dodoi{10.1038/s41586-020-2839-y}

\bibitem[{{Lorimer} {et~al.}(2007){Lorimer}, {Bailes}, {McLaughlin},
  {Narkevic}, \& {Crawford}}]{lorimer07}
{Lorimer}, D.~R., {Bailes}, M., {McLaughlin}, M.~A., {Narkevic}, D.~J., \&
  {Crawford}, F. 2007, Science, 318, 777, \dodoi{10.1126/science.1147532}

\bibitem[{{Lu} {et~al.}(2020){Lu}, {Kumar}, \& {Zhang}}]{lu20}
{Lu}, W., {Kumar}, P., \& {Zhang}, B. 2020, \mnras, 498, 1397,
  \dodoi{10.1093/mnras/staa2450}

\bibitem[{{Luo} {et~al.}(2025){Luo}, {Niu}, {Wang}, {Zhang}, {De-Jiang Zhou},
  {Xu}, {Wang}, {Niu}, {Zhang}, {Zhang}, {Cai}, {Han}, {Li}, {Lee}, {Zhu}, \&
  {Zhang}}]{luo25}
{Luo}, J.-W., {Niu}, J.-R., {Wang}, W.-Y., {et~al.} 2025, \apj, 988, 62,
  \dodoi{10.3847/1538-4357/ade0b9}

\bibitem[{{Luo} {et~al.}(2020){Luo}, {Men}, {Lee}, {Wang}, {Lorimer}, \&
  {Zhang}}]{luo20}
{Luo}, R., {Men}, Y., {Lee}, K., {et~al.} 2020, \mnras, 494, 665,
  \dodoi{10.1093/mnras/staa704}

\bibitem[{{Lyutikov}(2021)}]{lyutikov21}
{Lyutikov}, M. 2021, \apj, 922, 166, \dodoi{10.3847/1538-4357/ac1b32}

\bibitem[{{Lyutikov}(2024)}]{lyutikov24}
---. 2024, \mnras, 529, 2180, \dodoi{10.1093/mnras/stae591}

\bibitem[{{Margalit} {et~al.}(2020){Margalit}, {Beniamini}, {Sridhar}, \&
  {Metzger}}]{margalit20}
{Margalit}, B., {Beniamini}, P., {Sridhar}, N., \& {Metzger}, B.~D. 2020,
  \apjl, 899, L27, \dodoi{10.3847/2041-8213/abac57}

\bibitem[{{Margalit} {et~al.}(2019){Margalit}, {Berger}, \&
  {Metzger}}]{margalit19}
{Margalit}, B., {Berger}, E., \& {Metzger}, B.~D. 2019, \apj, 886, 110,
  \dodoi{10.3847/1538-4357/ab4c31}

\bibitem[{{Martin} {et~al.}(2014){Martin}, {Rea}, {Torres}, \&
  {Papitto}}]{martin14}
{Martin}, J., {Rea}, N., {Torres}, D.~F., \& {Papitto}, A. 2014, \mnras, 444,
  2910, \dodoi{10.1093/mnras/stu1594}

\bibitem[{{Mckinven} {et~al.}(2025){Mckinven}, {Bhardwaj}, {Eftekhari},
  {Kilpatrick}, {Kirichenko}, {Pal}, {Cook}, {Gaensler}, {Giri}, {Kaspi},
  {Michilli}, {Nimmo}, {Pearlman}, {Pleunis}, {Sand}, {Stairs}, {Andersen},
  {Andrew}, {Bandura}, {Brar}, {Cassanelli}, {Chatterjee}, {Curtin}, {Dong},
  {Eadie}, {Fonseca}, {Ibik}, {Kaczmarek}, {Kharel}, {Lazda}, {Leung}, {Li},
  {Main}, {Masui}, {Mena-Parra}, {Ng}, {Pandhi}, {Patil}, {Prochaska},
  {Rafiei-Ravandi}, {Scholz}, {Shah}, {Shin}, \& {Smith}}]{mckinven25}
{Mckinven}, R., {Bhardwaj}, M., {Eftekhari}, T., {et~al.} 2025, \nat, 637, 43,
  \dodoi{10.1038/s41586-024-08184-4}

\bibitem[{{Mereghetti} {et~al.}(2020){Mereghetti}, {Savchenko}, {Ferrigno},
  {G{\"o}tz}, {Rigoselli}, {Tiengo}, {Bazzano}, {Bozzo}, {Coleiro},
  {Courvoisier}, {Doyle}, {Goldwurm}, {Hanlon}, {Jourdain}, {von Kienlin},
  {Lutovinov}, {Martin-Carrillo}, {Molkov}, {Natalucci}, {Onori}, {Panessa},
  {Rodi}, {Rodriguez}, {S{\'a}nchez-Fern{\'a}ndez}, {Sunyaev}, \&
  {Ubertini}}]{Integral-SGR}
{Mereghetti}, S., {Savchenko}, V., {Ferrigno}, C., {et~al.} 2020, \apjl, 898,
  L29, \dodoi{10.3847/2041-8213/aba2cf}

\bibitem[{{Ng} \& {CHIME/FRB Collaboration}(2025)}]{ng25}
{Ng}, M., \& {CHIME/FRB Collaboration}. 2025, The Astronomer's Telegram, 17081,
  1

\bibitem[{{Nimmo} {et~al.}(2025){Nimmo}, {Pleunis}, {Beniamini}, {Kumar},
  {Lanman}, {Li}, {Main}, {Sammons}, {Andrew}, {Bhardwaj}, {Chatterjee},
  {Curtin}, {Fonseca}, {Gaensler}, {Joseph}, {Kader}, {Kaspi}, {Lazda},
  {Leung}, {Masui}, {Mckinven}, {Michilli}, {Pandhi}, {Pearlman},
  {Rafiei-Ravandi}, {Sand}, {Shin}, {Smith}, \& {Stairs}}]{nimmo25}
{Nimmo}, K., {Pleunis}, Z., {Beniamini}, P., {et~al.} 2025, \nat, 637, 48,
  \dodoi{10.1038/s41586-024-08297-w}

\bibitem[{{Niu} {et~al.}(2022{\natexlab{a}}){Niu}, {Aggarwal}, {Li}, {Zhang},
  {Chatterjee}, {Tsai}, {Yu}, {Law}, {Burke-Spolaor}, {Cordes}, {Zhang},
  {Ocker}, {Yao}, {Wang}, {Feng}, {Niino}, {Bochenek}, {Cruces}, {Connor},
  {Jiang}, {Dai}, {Luo}, {Li}, {Miao}, {Niu}, {Anna-Thomas}, {Sydnor}, {Stern},
  {Wang}, {Yuan}, {Yue}, {Zhou}, {Yan}, {Zhu}, \& {Zhang}}]{niu22}
{Niu}, C.~H., {Aggarwal}, K., {Li}, D., {et~al.} 2022{\natexlab{a}}, \nat, 606,
  873, \dodoi{10.1038/s41586-022-04755-5}

\bibitem[{{Niu} {et~al.}(2022{\natexlab{b}}){Niu}, {Zhu}, {Zhang}, {Yuan},
  {Zhou}, {Zhang}, {Jiang}, {Han}, {Li}, {Lee}, {Wang}, {Feng}, {Li}, {Luo},
  {Wang}, {Dai}, {Miao}, {Niu}, {Xu}, {Zhang}, {Wang}, {Wang}, \&
  {Xu}}]{niujr22}
{Niu}, J.-R., {Zhu}, W.-W., {Zhang}, B., {et~al.} 2022{\natexlab{b}}, Research
  in Astronomy and Astrophysics, 22, 124004, \dodoi{10.1088/1674-4527/ac995d}

\bibitem[{{Niu} {et~al.}(2024){Niu}, {Wang}, {Jiang}, {Qu}, {Zhou}, {Zhu},
  {Lee}, {Han}, {Zhang}, {Li}, {Cao}, {Fang}, {Feng}, {Fu}, {Jiang}, {Jing},
  {Li}, {Li}, {Luo}, {Meng}, {Miao}, {Miao}, {Niu}, {Pan}, {Wang}, {Wang},
  {Wang}, {Wang}, {Wu}, {Wu}, {Xu}, {Xu}, {Xu}, {Xue}, {Yang}, {Yuan}, {Yue},
  {Zhao}, {Zhang}, {Zhang}, {Zhang}, {Zhang}, {Zhang}, \& {Zhu}}]{niujr24}
{Niu}, J.~R., {Wang}, W.~Y., {Jiang}, J.~C., {et~al.} 2024, \apjl, 972, L20,
  \dodoi{10.3847/2041-8213/ad7023}

\bibitem[{{Olausen} \& {Kaspi}(2014)}]{Olausen2014}
{Olausen}, S.~A., \& {Kaspi}, V.~M. 2014, \apjs, 212, 6,
  \dodoi{10.1088/0067-0049/212/1/6}

\bibitem[{{Pal}(2025)}]{pal25}
{Pal}, A. 2025, \apjl, 983, L15, \dodoi{10.3847/2041-8213/adc386}

\bibitem[{{Petroff} {et~al.}(2019){Petroff}, {Hessels}, \&
  {Lorimer}}]{petroff19}
{Petroff}, E., {Hessels}, J.~W.~T., \& {Lorimer}, D.~R. 2019, \aapr, 27, 4,
  \dodoi{10.1007/s00159-019-0116-6}

\bibitem[{{Philippov} {et~al.}(2014){Philippov}, {Tchekhovskoy}, \&
  {Li}}]{Philippov2014}
{Philippov}, A., {Tchekhovskoy}, A., \& {Li}, J.~G. 2014, \mnras, 441, 1879,
  \dodoi{10.1093/mnras/stu591}

\bibitem[{{Platts} {et~al.}(2019){Platts}, {Weltman}, {Walters}, {Tendulkar},
  {Gordin}, \& {Kandhai}}]{platts19}
{Platts}, E., {Weltman}, A., {Walters}, A., {et~al.} 2019, \physrep, 821, 1,
  \dodoi{10.1016/j.physrep.2019.06.003}

\bibitem[{{Qian} {et~al.}(2025){Qian}, {Niu}, {Cao}, {Huang}, {Yang}, {Zhang},
  {Zhang}, {Zhu}, {Li}, {Pan}, {Sun}, {Zhang}, {Sun}, {Liu}, {Jin}, {Wu},
  {Yuan}, {Zhang}, {Zhu}, \& {Jiang}}]{qian25}
{Qian}, L., {Niu}, J.~R., {Cao}, J.~H., {et~al.} 2025, The Astronomer's
  Telegram, 17126, 1

\bibitem[{{Qu} {et~al.}(2022){Qu}, {Kumar}, \& {Zhang}}]{qu22b}
{Qu}, Y., {Kumar}, P., \& {Zhang}, B. 2022, \mnras, 515, 2020,
  \dodoi{10.1093/mnras/stac1910}

\bibitem[{{Qu} \& {Zhang}(2024)}]{quzhang24}
{Qu}, Y., \& {Zhang}, B. 2024, \apj, 972, 124, \dodoi{10.3847/1538-4357/ad5d5b}

\bibitem[{{Rajwade} {et~al.}(2020){Rajwade}, {Mickaliger}, {Stappers},
  {Morello}, {Agarwal}, {Bassa}, {Breton}, {Caleb}, {Karastergiou}, {Keane}, \&
  {Lorimer}}]{rajwade20}
{Rajwade}, K.~M., {Mickaliger}, M.~B., {Stappers}, B.~W., {et~al.} 2020,
  \mnras, \dodoi{10.1093/mnras/staa1237}

\bibitem[{{Ravi}(2019)}]{ravi19b}
{Ravi}, V. 2019, Nature Astronomy, 3, 928, \dodoi{10.1038/s41550-019-0831-y}

\bibitem[{{Reboul-Salze} {et~al.}(2025){Reboul-Salze}, {Barr{\`e}re}, {Kiuchi},
  {Guilet}, {Raynaud}, {Fujibayashi}, \& {Shibata}}]{Reboul-Salze2025}
{Reboul-Salze}, A., {Barr{\`e}re}, P., {Kiuchi}, K., {et~al.} 2025, \aap, 699,
  A4, \dodoi{10.1051/0004-6361/202453126}

\bibitem[{{Richardson} {et~al.}(2023){Richardson}, {Pavao}, {Eldridge},
  {Pablo}, {Chen{\'e}}, {Wysocki}, {Gies}, {Younes}, \& {Hare}}]{Richardson23}
{Richardson}, N.~D., {Pavao}, C.~M., {Eldridge}, J.~J., {et~al.} 2023, \nat,
  614, 45, \dodoi{10.1038/s41586-022-05618-9}

\bibitem[{{Schneider} {et~al.}(2019){Schneider}, {Ohlmann}, {Podsiadlowski},
  {R{\"o}pke}, {Balbus}, {Pakmor}, \& {Springel}}]{Schneider2019}
{Schneider}, F. R.~N., {Ohlmann}, S.~T., {Podsiadlowski}, P., {et~al.} 2019,
  \nat, 574, 211, \dodoi{10.1038/s41586-019-1621-5}

\bibitem[{{Scholz} {et~al.}(2017){Scholz}, {Bogdanov}, {Hessels}, {Lynch},
  {Spitler}, {Bassa}, {Bower}, {Burke-Spolaor}, {Butler}, {Chatterjee},
  {Cordes}, {Gourdji}, {Kaspi}, {Law}, {Marcote}, {McLaughlin}, {Michilli},
  {Paragi}, {Ransom}, {Seymour}, {Tendulkar}, \& {Wharton}}]{scholz17}
{Scholz}, P., {Bogdanov}, S., {Hessels}, J.~W.~T., {et~al.} 2017, \apj, 846,
  80, \dodoi{10.3847/1538-4357/aa8456}

\bibitem[{{Schwab}(2021)}]{Schwab2021}
{Schwab}, J. 2021, \apj, 906, 53, \dodoi{10.3847/1538-4357/abc87e}

\bibitem[{{Shenar} {et~al.}(2023){Shenar}, {Wade}, {Marchant}, {Bagnulo},
  {Bodensteiner}, {Bowman}, {Gilkis}, {Langer}, {Nicolas-Chen{\'e}},
  {Oskinova}, {Van Reeth}, {Sana}, {St-Louis}, {de Oliveira}, {Todt}, \&
  {Toonen}}]{Shenar2023}
{Shenar}, T., {Wade}, G.~A., {Marchant}, P., {et~al.} 2023, Science, 381, 761,
  \dodoi{10.1126/science.ade3293}

\bibitem[{{Sherman} {et~al.}(2024){Sherman}, {Ravi}, {El-Badry}, {Sharma},
  {Ocker}, {Kosogorov}, {Connor}, \& {Faber}}]{Sherman2024}
{Sherman}, M.~B., {Ravi}, V., {El-Badry}, K., {et~al.} 2024, \mnras, 531, 2379,
  \dodoi{10.1093/mnras/stae1289}

\bibitem[{{Sobacchi} {et~al.}(2024){Sobacchi}, {Iwamoto}, {Sironi}, \&
  {Piran}}]{sobacchi24}
{Sobacchi}, E., {Iwamoto}, M., {Sironi}, L., \& {Piran}, T. 2024, \aap, 690,
  A332, \dodoi{10.1051/0004-6361/202451725}

\bibitem[{{Spitkovsky}(2006)}]{Spitkovsky2006}
{Spitkovsky}, A. 2006, \apjl, 648, L51, \dodoi{10.1086/507518}

\bibitem[{{Spitler} {et~al.}(2016){Spitler}, {Scholz}, {Hessels}, {Bogdanov},
  {Brazier}, {Camilo}, {Chatterjee}, {Cordes}, {Crawford}, {Deneva}, {Ferdman},
  {Freire}, {Kaspi}, {Lazarus}, {Lynch}, {Madsen}, {McLaughlin}, {Patel},
  {Ransom}, {Seymour}, {Stairs}, {Stappers}, {van Leeuwen}, \&
  {Zhu}}]{spitler16}
{Spitler}, L.~G., {Scholz}, P., {Hessels}, J.~W.~T., {et~al.} 2016, \nat, 531,
  202, \dodoi{10.1038/nature17168}

\bibitem[{{Spruit}(2002)}]{spruit02}
{Spruit}, H.~C. 2002, \aap, 381, 923, \dodoi{10.1051/0004-6361:20011465}

\bibitem[{{Spruit}(2008)}]{spruit08}
{Spruit}, H.~C. 2008, in American Institute of Physics Conference Series, Vol.
  983, 40 Years of Pulsars: Millisecond Pulsars, Magnetars and More, ed.
  C.~{Bassa}, Z.~{Wang}, A.~{Cumming}, \& V.~M. {Kaspi} (AIP), 391--398,
  \dodoi{10.1063/1.2900262}

\bibitem[{{Sridhar} {et~al.}(2021){Sridhar}, {Metzger}, {Beniamini},
  {Margalit}, {Renzo}, {Sironi}, \& {Kovlakas}}]{sridhar21}
{Sridhar}, N., {Metzger}, B.~D., {Beniamini}, P., {et~al.} 2021, \apj, 917, 13,
  \dodoi{10.3847/1538-4357/ac0140}

\bibitem[{{Tayler}(1973)}]{taylor73}
{Tayler}, R.~J. 1973, \mnras, 161, 365, \dodoi{10.1093/mnras/161.4.365}

\bibitem[{{Thompson} \& {Duncan}(1993)}]{thompson93}
{Thompson}, C., \& {Duncan}, R.~C. 1993, \apj, 408, 194, \dodoi{10.1086/172580}

\bibitem[{{Thornton} {et~al.}(2013){Thornton}, {Stappers}, {Bailes},
  {Barsdell}, {Bates}, {Bhat}, {Burgay}, {Burke-Spolaor}, {Champion}, {Coster},
  {D'Amico}, {Jameson}, {Johnston}, {Keith}, {Kramer}, {Levin}, {Milia}, {Ng},
  {Possenti}, \& {van Straten}}]{thornton13}
{Thornton}, D., {Stappers}, B., {Bailes}, M., {et~al.} 2013, Science, 341, 53,
  \dodoi{10.1126/science.1236789}

\bibitem[{{Vink} \& {Kuiper}(2006)}]{vink06}
{Vink}, J., \& {Kuiper}, L. 2006, \mnras, 370, L14,
  \dodoi{10.1111/j.1745-3933.2006.00178.x}

\bibitem[{{Wada} {et~al.}(2021){Wada}, {Ioka}, \& {Zhang}}]{wada21}
{Wada}, T., {Ioka}, K., \& {Zhang}, B. 2021, \apj, 920, 54,
  \dodoi{10.3847/1538-4357/ac127a}

\bibitem[{{Wang} {et~al.}(2020){Wang}, {Wang}, {Yang}, {Yu}, {Zuo}, \&
  {Dai}}]{wang20}
{Wang}, F.~Y., {Wang}, Y.~Y., {Yang}, Y.-P., {et~al.} 2020, \apj, 891, 72,
  \dodoi{10.3847/1538-4357/ab74d0}

\bibitem[{{Wang} {et~al.}(2022){Wang}, {Zhang}, {Dai}, \& {Cheng}}]{wangfy22}
{Wang}, F.~Y., {Zhang}, G.~Q., {Dai}, Z.~G., \& {Cheng}, K.~S. 2022, Nature
  Communications, 13, 4382, \dodoi{10.1038/s41467-022-31923-y}

\bibitem[{{Wang} {et~al.}(2019){Wang}, {Zhang}, {Chen}, \& {Xu}}]{wang19}
{Wang}, W., {Zhang}, B., {Chen}, X., \& {Xu}, R. 2019, \apjl, 876, L15,
  \dodoi{10.3847/2041-8213/ab1aab}

\bibitem[{{Wang} {et~al.}(2024){Wang}, {Zhang}, {Zhou}, {Liu}, {Niu}, {Zhou},
  {Gao}, {Liu}, {Xu}, \& {Zhang}}]{wangwy24}
{Wang}, W.-Y., {Zhang}, C., {Zhou}, E., {et~al.} 2024, Research in Astronomy
  and Astrophysics, 24, 105012, \dodoi{10.1088/1674-4527/ad74db}

\bibitem[{{Xu} {et~al.}(2022){Xu}, {Niu}, {Chen}, {Lee}, {Zhu}, {Dong},
  {Zhang}, {Jiang}, {Wang}, {Xu}, {Zhang}, {Fu}, {Filippenko}, {Peng}, {Zhou},
  {Zhang}, {Wang}, {Feng}, {Li}, {Brink}, {Li}, {Lu}, {Yang}, {Caballero},
  {Cai}, {Chen}, {Dai}, {Djorgovski}, {Esamdin}, {Gan}, {Guhathakurta}, {Han},
  {Hao}, {Huang}, {Jiang}, {Li}, {Li}, {Li}, {Li}, {Li}, {Liu}, {Luo}, {Men},
  {Niu}, {Peng}, {Qian}, {Song}, {Stern}, {Stockton}, {Sun}, {Wang}, {Wang},
  {Wang}, {Wang}, {Wu}, {Xiao}, {Xiong}, {Xu}, {Xu}, {Yang}, {Yang}, {Yao},
  {Yi}, {Yue}, {Yu}, {Yu}, {Yuan}, {Zhang}, {Zhang}, {Zhang}, {Zhao}, {Zheng},
  {Zhu}, \& {Zou}}]{xuh22}
{Xu}, H., {Niu}, J.~R., {Chen}, P., {et~al.} 2022, \nat, 609, 685,
  \dodoi{10.1038/s41586-022-05071-8}

\bibitem[{{Xu} {et~al.}(2025){Xu}, {Xu}, {Guo}, {Jiang}, {Wang}, {Xue}, {Men},
  {Lee}, {Zhang}, {Zhu}, \& {Han}}]{xujw25}
{Xu}, J., {Xu}, H., {Guo}, Y., {et~al.} 2025, arXiv e-prints, arXiv:2505.06006,
  \dodoi{10.48550/arXiv.2505.06006}

\bibitem[{{Yang} {et~al.}(2022){Yang}, {Ai}, {Zhang}, {Zhang}, {Liu}, {Wang},
  {Yang}, {Yin}, {Li}, \& {L{\"u}}}]{yangj22}
{Yang}, J., {Ai}, S., {Zhang}, B.-B., {et~al.} 2022, \nat, 612, 232,
  \dodoi{10.1038/s41586-022-05403-8}

\bibitem[{{Yang} {et~al.}(2023){Yang}, {Xu}, \& {Zhang}}]{yang23b}
{Yang}, Y.-P., {Xu}, S., \& {Zhang}, B. 2023, \mnras, 520, 2039,
  \dodoi{10.1093/mnras/stad168}

\bibitem[{{Yang} \& {Zhang}(2018)}]{yangzhang18}
{Yang}, Y.-P., \& {Zhang}, B. 2018, \apj, 868, 31,
  \dodoi{10.3847/1538-4357/aae685}

\bibitem[{{Yang} \& {Zhang}(2021)}]{yangzhang21}
---. 2021, \apj, 919, 89, \dodoi{10.3847/1538-4357/ac14b5}

\bibitem[{{Yu} {et~al.}(2013){Yu}, {Zhang}, \& {Gao}}]{yu13}
{Yu}, Y.-W., {Zhang}, B., \& {Gao}, H. 2013, \apjl, 776, L40,
  \dodoi{10.1088/2041-8205/776/2/L40}

\bibitem[{{Zhang}(2013)}]{zhang13}
{Zhang}, B. 2013, \apjl, 763, L22, \dodoi{10.1088/2041-8205/763/1/L22}

\bibitem[{{Zhang}(2017)}]{zhang17}
---. 2017, \apjl, 836, L32, \dodoi{10.3847/2041-8213/aa5ded}

\bibitem[{{Zhang}(2018)}]{zhang18b}
---. 2018, \apjl, 854, L21, \dodoi{10.3847/2041-8213/aaadba}

\bibitem[{{Zhang}(2020)}]{zhang20b}
---. 2020, \nat, 587, 45, \dodoi{10.1038/s41586-020-2828-1}

\bibitem[{{Zhang}(2022)}]{zhang22}
---. 2022, \apj, 925, 53, \dodoi{10.3847/1538-4357/ac3979}

\bibitem[{{Zhang}(2023)}]{zhang23}
---. 2023, Reviews of Modern Physics, 95, 035005,
  \dodoi{10.1103/RevModPhys.95.035005}

\bibitem[{{Zhang} {et~al.}(2025){Zhang}, {Wang}, {Wang}, {Gao}, {Wu}, {Li},
  {Zhu}, {Zhang}, {Lee}, {Han}, {Tsai}, {Wang}, {Huang}, {Zou}, {Zhou}, {Lu},
  {Xie}, {Fang}, {Cao}, {Miao}, {Zhu}, {Chen}, {Cheng}, {Ke}, {Zhang}, {Zhang},
  {Cao}, {Tian}, {Wu}, {Zhang}, {Niu}, {Zhou}, {Xu}, {Wang}, {Chen}, {Chen},
  {Cui}, {Feng}, {G{\"u}gercino{\u{g}}lu}, {Huang}, {Li}, {Li}, {Li}, {Li},
  {Lin}, {Liu}, {Luo}, {Luo}, {Niu}, {Qu}, {Qu}, {Menberu Tedila}, {Wang},
  {Wang}, {Wang}, {Wang}, {Weng}, {Wu}, {Xu}, {Yang}, {Yang}, {Yew}, {Yu},
  {Zhang}, \& {Zhao}}]{zhangjs25}
{Zhang}, J.-S., {Wang}, T.-C., {Wang}, P., {et~al.} 2025, arXiv e-prints,
  arXiv:2507.14707, \dodoi{10.48550/arXiv.2507.14707}

\bibitem[{{Zhang} \& {Gao}(2020)}]{zhanggao20}
{Zhang}, X., \& {Gao}, H. 2020, \mnras, 498, L1, \dodoi{10.1093/mnrasl/slaa116}

\bibitem[{{Zhang} {et~al.}(2018){Zhang}, {Gajjar}, {Foster}, {Siemion},
  {Cordes}, {Law}, \& {Wang}}]{zhangy18}
{Zhang}, Y.~G., {Gajjar}, V., {Foster}, G., {et~al.} 2018, \apj, 866, 149,
  \dodoi{10.3847/1538-4357/aadf31}

\bibitem[{{Zhang} {et~al.}(2022){Zhang}, {Wang}, {Feng}, {Zhang}, {Li}, {Tsai},
  {Niu}, {Luo}, {Yao}, {Zhu}, {Han}, {Lee}, {Zhou}, {Niu}, {Jiang}, {Wang},
  {Zhang}, {Xu}, {Wang}, \& {Xu}}]{zhangyk22}
{Zhang}, Y.-K., {Wang}, P., {Feng}, Y., {et~al.} 2022, Research in Astronomy
  and Astrophysics, 22, 124002, \dodoi{10.1088/1674-4527/ac98f7}

\bibitem[{{Zhang} {et~al.}(2023){Zhang}, {Li}, {Zhang}, {Cao}, {Feng}, {Wang},
  {Qu}, {Niu}, {Zhu}, {Han}, {Jiang}, {Lee}, {Li}, {Luo}, {Niu}, {Tsai},
  {Wang}, {Wang}, {Wu}, {Xu}, {Yang}, {Zhang}, {Zhou}, \& {Zhu}}]{zhangyk23}
{Zhang}, Y.-K., {Li}, D., {Zhang}, B., {et~al.} 2023, \apj, 955, 142,
  \dodoi{10.3847/1538-4357/aced0b}

\bibitem[{{Zhong} \& {Dai}(2020)}]{Zhong2020}
{Zhong}, S.-Q., \& {Dai}, Z.-G. 2020, \apj, 893, 9,
  \dodoi{10.3847/1538-4357/ab7bdf}

\bibitem[{{Zhou} {et~al.}(2025{\natexlab{a}}){Zhou}, {Han}, {Zhang}, {Zhu},
  {Wang}, {Yang}, {Qu}, {Zhang}, {Yan}, {Jing}, {Cao}, {Xie}, {Yang}, {Tian},
  {Li}, {Li}, {Niu}, {Wu}, {Wu}, {Feng}, {Wang}, \& {Wang}}]{zhou25}
{Zhou}, D., {Han}, J.~L., {Zhang}, B., {et~al.} 2025{\natexlab{a}}, \apj, 988,
  41, \dodoi{10.3847/1538-4357/addfdb}

\bibitem[{{Zhou} {et~al.}(2025{\natexlab{b}}){Zhou}, {Wang}, {Fang}, {Zhu},
  {Zhang}, {Li}, {Feng}, {Huang}, {Lee}, {Han}, {Zou}, {Zhang}, {Luo}, {Zhang},
  {Wang}, {Lu}, {Cao}, {Yu}, {Li}, {Miao}, {Xie}, {Chen}, {Qu}, {Chen}, {Zhu},
  {Cao}, {Chen}, {Du}, {Gao}, {Huang}, {Li}, {Li}, {Li}, {Lin}, {Liu}, {Luo},
  {Niu}, {Niu}, {Qu}, {Tian}, {Tsai}, {Wang}, {Wang}, {Wang}, {Wang}, {Weng},
  {Wu}, {Wu}, {Xu}, {Yang}, {Yew}, {Zhang}, {Zhang}, {Zhang}, {Zhao}, \&
  {Zhou}}]{zhoudk25}
{Zhou}, D., {Wang}, P., {Fang}, J., {et~al.} 2025{\natexlab{b}}, arXiv
  e-prints, arXiv:2507.14708, \dodoi{10.48550/arXiv.2507.14708}

\bibitem[{{Zhou} {et~al.}(2022){Zhou}, {Han}, {Zhang}, {Lee}, {Zhu}, {Li},
  {Jing}, {Wang}, {Zhang}, {Jiang}, {Niu}, {Luo}, {Xu}, {Zhang}, {Wang}, {Xu},
  {Wang}, {Yang}, \& {Feng}}]{zhou22}
{Zhou}, D.~J., {Han}, J.~L., {Zhang}, B., {et~al.} 2022, Research in Astronomy
  and Astrophysics, 22, 124001, \dodoi{10.1088/1674-4527/ac98f8}

\bibitem[{{Zhu} {et~al.}(2023){Zhu}, {Xu}, {Zhou}, {Lin}, {Wang}, {Wang},
  {Zhang}, {Niu}, {Chen}, {Li}, {Meng}, {Lee}, {Zhang}, {Feng}, {Ge},
  {G{\"o}{\u{g}}{\"u}{\c{s}}}, {Guan}, {Han}, {Jiang}, {Jiang}, {Kouveliotou},
  {Li}, {Miao}, {Miao}, {Men}, {Niu}, {Wang}, {Wang}, {Xu}, {Xu}, {Xue},
  {Yang}, {Yu}, {Yuan}, {Yue}, {Zhang}, \& {Zhang}}]{zhu23}
{Zhu}, W., {Xu}, H., {Zhou}, D., {et~al.} 2023, Science Advances, 9, eadf6198,
  \dodoi{10.1126/sciadv.adf6198}

\end{thebibliography}

\end{document}